\documentclass[a4paper,11pt]{article}
\pdfoutput=1
\usepackage[colorlinks=true]{hyperref}
\usepackage[dvipsnames]{xcolor}
\usepackage{graphicx,rotating,hyperref,slashed,amsmath,xcolor,amssymb,amsfonts,colortbl,cite}
\usepackage{microtype}
\makeatletter
\hypersetup{colorlinks,bookmarksopen,bookmarksnumbered,  
linkcolor=blus,pdfstartview=FitH,urlcolor=rossos,citecolor=verde}
\allowdisplaybreaks		 

\hypersetup{%
    urlcolor=black,
    citecolor=black,
    linkcolor=black
}
\usepackage{mciteplus}

\AtBeginDocument{
  \hypersetup{
     urlcolor=NavyBlue,
    citecolor=NavyBlue,
    linkcolor=black,
  }%
}
\newcommand{\oo}[1]{\textcolor{brown}{[{\bf OO}: #1]}}
\newcommand{\gt}[1]{\textcolor{blue}{#1}}

\usepackage{colortbl}
\definecolor{green3}{cmyk}{0.8, 0., 0.7., 0.}
\definecolor{green2}{cmyk}{0, 1, 0.5, 0}
\definecolor{lightgreen}{cmyk}{0.2, 0, 0.2, 0.2}
\definecolor{lightgray}{cmyk}{0.1,0.2,0,0.1}
\definecolor{lightgray2}{cmyk}{0.4,0.4,0,0.8}
\definecolor{black}{cmyk}{1.0,1.0,1.0,1.0}

\allowdisplaybreaks[1]

\usepackage{colortbl}
\definecolor{lightgreen}{cmyk}{0.2, 0, 0.2, 0.2}
\definecolor{lightgray}{cmyk}{0.1,0.2,0,0.1}
\definecolor{lightgray2}{cmyk}{0.1,0.1,0,0.1}

\makeatletter
\newlength{\apb@width}
\newcommand{\autoparbox}[2][c]{\settowidth{\apb@width}{#2}\parbox[#1]{\apb@width}{#2}}

\makeatother

\numberwithin{equation}{section}

\def\beq{\begin{equation}}
\def\eeq{\end{equation}}

\def\gt{\color{blue}}
\def\bea{\begin{eqnarray}}
\def\eea{\end{eqnarray}}
\def\eg{{\it e.g.~}}

\def\ie{{\it i.e.~}}
\def\d{{\rm d}}

\def\d{{\rm d}}

\def\nn{\nonumber}

\def\Mp{M_{\rm pl}}

\def\l{{\ell}}

\def\fr{\frac}

\def\0{{\boldsymbol 0}}

\def\x{{\boldsymbol{x}}}

\def\fr{\frac}

\def\etc{\eta_{\rm c}}
\def\dn{\Delta N}

\def\cc{\mathcal{C}}

\usepackage{setspace} 

\begin{document}

\begin{titlepage}

\setcounter{page}{1} \baselineskip=15.5pt \thispagestyle{empty}

\bigskip\

\vspace{1cm}
\begin{center}

{
{\fontsize{19}{28}\selectfont  \sffamily \bfseries {
{CMB $\mu T$ cross-correlations as a probe of  PBH scenarios}
}}
}
\\
\hskip1cm

\end{center}

\vspace{0.2cm}

\begin{center}
{\fontsize{13}{30}\selectfont Ogan \"Ozsoy$^{\heartsuit}$
\,, Gianmassimo Tasinato$^{\spadesuit}$}
\end{center}

\begin{center}

\vskip 8pt
\textsl{
$\spadesuit$ Department of Physics, Swansea University, Swansea, SA2 8PP, United Kingdom\\
$\heartsuit$ CEICO, Institute of Physics of the Czech Academy of Sciences, Na Slovance 1999/2, 182 21,
Prague.}
\vskip 7pt

\end{center}

\vspace{1.2cm}
\hrule \vspace{0.3cm}
\noindent {\sffamily \bfseries Abstract} \\[0.1cm]    
  We propose a new method for probing inflationary models of primordial black hole (PBH) production, using only CMB physics at relatively large
scales. In PBH scenarios,
   the primordial power spectrum profile for curvature perturbations is characterized by a pronounced
   dip, followed by a rapid growth towards small scales, leading to a peak  responsible for 
  PBH formation. We focus on scales around the dip that are well separated from the peak to analytically compute expressions
  for the curvature power spectrum and bispectrum. 
 The size of the squeezed  bispectrum is  enhanced at the  position of the dip, and  it  acquires a characteristic scale dependence that  can  be  probed by cross-correlating
 CMB $\mu$-distortions and  temperature fluctuations. We quantitatively  study
 the properties of such cross-correlations  and how they depend on the underlying model, discussing how
 they can be tested by the next generation of CMB $\mu$-distortion experiments. This
 method allows one to experimentally probe inflationary PBH scenarios using well-understood CMB physics, without considering  non-linearities
  associated with   PBH formation  and  evolution.
\vskip 10pt
\hrule

\vspace{0.6cm}
 \end{titlepage}

 \tableofcontents
 
\newpage

\section{Introduction}
Gravitational waves from merging black holes, in conjunction with 
other  probes, can be used for testing our understanding
of cosmology. There is the intriguing possibility  that part of black holes from merging
events have primordial origin, arising from the collapse of
overdense regions in the early stages of our universe evolution \cite{Hawking:1971ei,Carr:1974nx,Carr:1975qj}. 
The existence of such overdense regions  might be attributed
to the dynamics of inflationary cosmology\footnote{See \cite{Khlopov:2008qy,Belotsky:2014kca} for alternative mechanisms of PBH production and their implications.} \cite{Ivanov:1994pa,GarciaBellido:1996qt}.  In order for producing
primordial black holes (PBH) with    astrophysically relevant masses, the spectrum of primordial curvature
perturbations at small scales should increase by several orders
of magnitude with respect to its values at large, CMB scales.  Within the context of single-field inflation, such 
an amplification can be realized  during a short non-attractor period, when the would-be decaying mode  influences the evolution
of super-horizon fluctuations. In this set-up, a peak in the curvature perturbation
spectrum can be produced, and the non-linear process of PBH production and subsequent
evolution depends  on the details of the peak, as first explicitly discussed in \cite{Germani:2018jgr}. 
 We refer the reader to the  
reviews \cite{Carr:2016drx,Sasaki:2018dmp,Carr:2020xqk,Green:2020jor} for details and  references on these  topics. 

In this work, we ask whether we can  probe the PBH formation mechanism by
testing larger scales well away from the peak of the spectrum. In fact, a close examination
of the profile of the spectrum in inflationary scenarios capable of producing PBHs shows universal features,
that are interesting to investigate. In particular, the spectrum is typically characterized by a very pronounced dip, followed
by a rapid growth with a well defined slope towards its peak  (see e.g. \cite{Byrnes:2018txb}).    The position of the dip is not random, but  depends  on other global properties  of the spectrum, as well as on details of the
non-attractor inflationary evolution \cite{Tasinato:2020vdk}. It typically occurs at much larger scales than the peak, $k_{\rm dip}/k_{\rm peak}\,\simeq\,10^{-2}$. Since such a dip seems to be a rather universal feature of power spectrum in single-field
models of PBH formation, probing further the physics associated with it offers an indirect way of testing these models, independently on
the  details of non-linear PBH formation and evolution mechanisms, at scales much larger than $k_{\rm peak}$. 

We start our analysis in Section \ref{SecHeu}, by analytically computing some of the key  properties of the power spectrum and bispectrum of curvature
fluctuations around the dip. For this purpose, we make use of the gradient expansion formalism introduced in \cite{Leach:2001zf},
which allows one to compute the evolution of fluctuations on super-horizon scales in models
with non-attractor phases \cite{Ozsoy:2019lyy}. Among our  new findings,   we show that the size of the bispectrum is generally amplified
at the scale of the dip position. It has a rich dependence on momenta, with a broad support spanning  different bispectrum  shapes.  When focussing
on isosceles and squeezed configurations,  it  acquires a characteristic momentum dependence
that is controlled by  the underlying  inflationary mechanism.

In Section \ref{sec-pheno}, we then propose a potential probe of such a scale dependent bispectrum around the dip position by utilizing cross-correlations between
CMB $\mu$-type distortions and temperature anisotropies at large scales. In fact,  $\mu T$ cross correlations are known to be 
sensitive to primordial non-Gaussianity \cite{Pajer:2012vz}, hence they represent  an  appropriate observable
for testing the features associated with it around the location of the dip. We show quantitatively  how such cross-correlations are influenced by the characteristic   scale-dependent   bispectrum in isosceles configurations, and  we estimate how our  results depend on the underlying
non-attractor evolution during inflation.  We also discuss  the prospects for detectability of this signal with
future $\mu$-distortion experiments, showing that  $\mu T$ cross-correlations constitute a promising avenue
to test PBH scenarios with clean and well-understood CMB physics only. In Section \ref{sec-dis}  we conclude with future directions, and we then include four technical appendixes.

\section{Curvature perturbation at super-horizon scales}
\label{SecHeu}

Models of primordial black hole (PBH) formation based on single field inflation
often include a short phase of non-attractor evolution, characterized by a transient growth of would be the decaying mode
which influences the super-horizon evolution of cosmological scalar perturbations\footnote{Direct enhancement of tensor perturbations on super-horizon scales can be also achieved by devising an analogue non-attractor phase within the generalized scalar-tensor theories of single field inflation \cite{Mylova:2018yap,Ozsoy:2019slf}.}.  In this section, we briefly describe a convenient formalism to study the statistics of curvature fluctuations in such situations. We aim to show that both the power
spectrum and the bispectrum are characterized by  pronounced  features, whose phenomenological   implications
are then  analyzed in Section \ref{sec-pheno}.

We assume  an isotropic, conformally
flat FLRW background metric
\begin{equation}
\d s^2\,=\,a^2(\tau) \left(- \d \tau^2+\d \vec x^2 \right)\,,
\end{equation}
where $a(\tau)$ is the scale factor, connecting conformal to physical time through the relation $a(\tau)\,\d \tau\,=\,\d t$.
Within canonical single-field inflation,
the dynamics of the curvature perturbation on comoving hypersurfaces, $\mathcal{R}_k (\tau)$,  is governed by the following equation in Fourier space \cite{Mukhanov:2005sc}
\beq\label{CPE}
\frac{1}{z^{2}(\tau)}\left[z^{2}(\tau) \mathcal{R}_{k}^{\prime}(\tau)\right]^{\prime}=-k^{2} \mathcal{R}_{k}(\tau)
\eeq
where $z\,=\,a \,\dot \phi/H$ is the so-called pump field, with $\phi$ the background inflaton profile and $H\,=\,\dot a/a$ the Hubble parameter. Primes indicate derivatives along conformal time, and dots derivatives along physical time.

To study the dynamics of fluctuations at super-horizon scales ($k \to 0$) we focus on the {\it gradient expansion formalism} introduced in  \cite{Leach:2001zf}.  We refer
the reader to \cite{Ozsoy:2019lyy} for further developments and applications of this formalism to the analysis of power spectrum of scalar fluctuations that leads to PBH formation. 
In this framework, iterative solutions of \eqref{CPE} can be generated at a desired order in $k\to 0$ expansion in terms of simple analytic functions describing the background evolution which then allow us to relate the late time $\tau = \tau_*$ (\ie at the reheating surface) curvature perturbation $\mathcal{R}_k(\tau)$ during inflation, to its value at an initial time around horizon exit $\tau = \tau_k$ in terms of a complex, $k$ dependent coefficient:
 \beq\label{irf}
\mathcal{R}_k (\tau_*) = \alpha_k \mathcal{R}_k (\tau_k)\,.
\eeq
Once expanded up to second order, $k^2$ gradient expansion,
  the  coefficient $\alpha_k$ reads
\beq\label{defalk}
\alpha_k = 1 + D(\tau_k)\, v_\mathcal{R} - F(\tau_k)\, k^2 +\mathcal{O}(k^4)\,,
\eeq
where we defined $k$ dependent fractional velocity of the curvature perturbation as 
\beq\label{fr}
v_\mathcal{R} (\tau_k) = \fr{\mathcal{R}_k'}{3\mathcal{H}_k\mathcal{R}_k} \bigg|_{\tau=\tau_k} \,.
\eeq 
The full $k$ dependence of the expression \eqref{defalk} on super-horizon scales is then encoded in $v_\mathcal{R}$ (See Appendix \ref{AppA}) and  the functions $D(\tau_k),F(\tau_k)$ which are given by the following nested integrals of the pump field $z(\tau) = a \dot{\phi}/H$ (See \cite{Leach:2001zf,Ozsoy:2019lyy} for further details):
\begin{align}
\label{Dint} D(\tau)&=3 {\cal H}_k\,\int_\tau^{\tau_*}\,\d \tau'\,\frac{z^2 (\tau_k) }{z^2 (\tau') }\, ,\\ 
\label{Fint}  F(\tau)&=\int_\tau^{\tau_*}\,\frac{d \tau'}{z^2(\tau')}
\,\int^{\tau'}_{\tau_k}\,\d \tau'' z^2(\tau'') \,. 
\end{align}
Whenever the pump field increases with time -- as in standard slow-roll inflation, where $z \propto a(\tau)$ --  the functions $D$, $F$ rapidly decrease to zero after
horizon crossing (\ie $\alpha_k \to 1$), and the curvature perturbation in \eqref{irf} settles to a constant shortly after horizon exit ($\mathcal{R}_k (\tau_*) \simeq \mathcal{R}_k (\tau_k)$). On the contrary, in inflationary models containing phases of non-attractor evolution, $z(\tau)$ transiently decreases  and the  functions $D$, $F$ can grow and amplify the curvature perturbation (\ie $|\alpha_k| \gg 1$ in \eqref{irf}) at super-horizon scales.
 
Making use of eq \eqref{irf},  the power spectrum for the curvature fluctuation $\mathcal{R}$ at late times can be related to the power spectrum evaluated at the horizon crossing via
\beq\label{psf1}
\mathcal{P}_\mathcal{R}(\tau_*,k) \equiv \fr{k^3}{2\pi^2}\langle\mathcal{R}_k(\tau_*)\mathcal{R}_{k'}(\tau_*)\rangle = |\alpha_k|^2 \, \mathcal{P}_\mathcal{R}(\tau_k)\,\delta\left(\vec{k}+\vec{k}'\right) ,
\eeq
where $\mathcal{P}_{\mathcal{R}}(\tau_k) \equiv k^3 |\mathcal{R}(\tau_k)|^2/2\pi^2$, $|\alpha_k|^2 = (\alpha_k^{R})^2 + (\alpha_k^{I})^2$, and we split $\alpha_k$ into its real and imaginary parts using $v_{\mathcal{R}}=v_\mathcal{R}^R + i \,v_\mathcal{R}^I$. Up to order $\mathcal{O}(k^2)$ in the gradient expansion, the real and the imaginary part of the enhancement factor are therefore given by
\begin{align}\label{ar}
\alpha_k^R &= 1 + D(\tau_k)\, v_\mathcal{R}^{R} - F(\tau_k)\, k^2,\\
\alpha_k^I &= D(\tau_k)\, v_{\mathcal{R}}^I. \label{ai}
\end{align}
We assume that $\mathcal{R}_k(\tau_k)$ is a Gaussian random variable: nevertheless, the superhorizon evolution typically introduces
non-linearities. In fact, we can go beyond the linear theory provided in eq. \eqref{irf} to compute 
the bispectrum of the late time curvature perturbation $\mathcal{R}_k(\tau_*)$. For the purpose of deriving an analytic expression for the bispectrum, we adopt the following non-linear version
for the curvature perturbation,   derived in \cite{Takamizu:2010xy} 
\beq\label{RNL}
\mathcal{R}_k(\tau_*)=\alpha_k \mathcal{R}_{k}(\tau_{k})+\fr{F(\tau_k)}{2}\left\{\int \frac{\d^{3} k^{\prime} \d^{3} k^{\prime \prime}}{(2 \pi)^{3}}\left(4 k^{\prime 2}-\delta_{i j} k^{\prime i} k^{\prime \prime j}\right) \mathcal{R}_{k'}(\tau_{k'}) \mathcal{R}_{k''}(\tau_{k''}) \delta\left(-\vec{k}+\vec{k}'+\vec{k}''\right)\right\}\,.
\eeq
where the last term represents the non-linear contribution parametrized by the convolution of the Gaussian variable $\mathcal{R}_k(\tau_k)$. Defining the corresponding bispectrum as 
\beq\label{BSdef}
\left\langle\mathcal{R}_{{k}_{1}}(\tau_*) \mathcal{R}_{{k}_{2}}(\tau_*) \mathcal{R}_{{k}_{3}}(\tau_*) \right\rangle=(2 \pi)^{3} B_{\mathcal{R}}\left({k}_{1}, {k}_{2}, {k}_{3}\right) \delta\left(\vec{k}_{1}+\vec{k}_{2}+\vec{k}_{3}\right),
\eeq
and using \eqref{RNL}, the bispectrum results with \cite{Takamizu:2010xy}
\beq\label{3PT}
B_{\mathcal{R}}\left({k}_{1}, {k}_{2}, {k}_{3}\right)=\frac{(2 \pi^{2})^2}{2(k_{1} k_{2} k_{3})^{3}}\left[\alpha^{*}_{k_{1}} \alpha_{k_{2}} F(\tau_{k_{3}})\left\{5\left(k_{1}^{2}+k_{2}^{2}\right)-k_{3}^{2}\right\} k_{3}^{3}\,\mathcal{P}_{\mathcal{R}}(\tau_{k_{1}}) \mathcal{P}_{\mathcal{R}}(\tau_{k_{2}})+\text { perms }\right],
\eeq
where permutations are among the three external wave-numbers. Correspondingly, we define the scale-dependent non-linearity parameter $f_{\rm NL}$ as
\beq\label{deffnl}
 f_{\rm NL}\left(k_{1}, k_{2}, k_{3}\right) =  \frac{5}{6} \fr{B_{\mathcal{R}}\left(k_{1}, k_{2}, k_{3}\right)}{\left[P_{\mathcal{R}}\left(\tau_*,k_1\right) P_{\mathcal{R}}\left(\tau_*,k_2 \right)+\text { perms }\right]}\,\,,
\eeq
where $P_{\mathcal{R}}(\tau_*,k)\equiv \frac{2\pi^2}{k^{3}} \mathcal{P}_{\mathcal{R}}(\tau_*,k)$. Using \eqref{psf1}, we then have
\beq\label{fnl}
 f_{\rm NL}\left(k_{1}, k_{2}, k_{3}\right) =\fr{5(k_1k_2k_3)^3}{24\pi^4} \fr{B_{\mathcal{R}}\left(k_{1}, k_{2}, k_{3}\right)}{\left[\left|\alpha_{k_1}\alpha_{k_2}\right|^2 \mathcal{P}_{\mathcal{R}}\left(\tau_{k_1}\right) \mathcal{P}_{\mathcal{R}}\left(\tau_{k_2} \right)k_3^3+\text { perms }\right]}.
\eeq
Finally, plugging the bispectrum in \eqref{3PT} into \eqref{fnl}, we re-write the scale-dependent 
 $f_{\rm NL}$ as
\beq\label{fnlf}
 f_{\rm NL}\left(k_{1}, k_{2}, k_{3}\right) = \fr{5}{12}\fr{\left(\alpha^{*}_{k_{1}} \alpha_{k_{2}} F(\tau_{k_{3}})\left\{5\left(k_{1}^{2}+k_{2}^{2}\right)-k_{3}^{2}\right\} k_{3}^{3}\,\mathcal{P}_{\mathcal{R}}(\tau_{k_{1}}) \mathcal{P}_{\mathcal{R}}(\tau_{k_{2}})+\text { perms }\right)}{\left[\left|\alpha_{k_1}\alpha_{k_2}\right|^2 \mathcal{P}_{\mathcal{R}}\left(\tau_{k_1}\right) \mathcal{P}_{\mathcal{R}}\left(\tau_{k_2} \right)k_3^3+\text { perms }\right]}.
\eeq
Notice from \eqref{fnlf} that the size and the scale-dependence of the $f_{\rm NL}$ parameter depends on the function $ F(\tau_{k})$ and the quantity $\alpha_k$, 
whose behavior depend on the background dynamics during inflation and in particular for the case of interest on the properties of the non-attractor regime. As a consequence, the corresponding bispectrum shape and its
scale-dependence can be very rich, as we will learn in what follows.

Armed with these preliminary, general results, we now study the scale dependence of the power spectrum
and bispectrum on a representative set-up capable of producing a large PBH population during inflation.

\subsection{The spectral shape of the  spectrum: a dip is followed by a rapid growth}\label{S2p3}

In order to study the evolution and enhancement in the power spectrum, we consider a representative  scenario that instantly connects an initial slow-roll era, with $\eta_{\rm sr} =0$,  to a slow-roll violating, non-attractor phase with  constant $\etc \leq -6$ where $\eta$ denotes the second slow-roll parameter, $H\,\eta\,\equiv\,d \ln \epsilon/d t$ and the first  slow-roll parameter is given by $H\,\epsilon\,=\,-d \ln H/d t$. In this setup, the pump field $z(\tau)$ is assumed to have a profile:
\beq\label{zsol1}
z(\tau)=
\left\{    
 \begin{array}{rl}
&z_0\left({\tau}/{\tau_0}\right)^{-1} \hskip1.6cm \,\,{\tau}/{\tau_0}\, \geq\, 1 \,,\\
&z_0 \left({\tau}/{\tau_0}\right)^{-(\eta_{\rm c}+2)/2} \hskip0.6cm\,\, {\tau_f}/{\tau_0}\leq {\tau}/{\tau_0} \leq 1 \,,
\end{array}\right. \,
\eeq 
{describing collectively the slow-roll and the constant-roll phases, the latter being controlled by the negative $\mathcal{O}(1)$ ‘‘slow-roll" parameter $\eta_c$.
We define $\tau_0$ as the transition time to the constant-roll era, $\tau_f$ as the conformal time when the constant-roll era ends, while we relate the quantity $z_0$ with a constant slow-roll parameter  $\epsilon_{\rm sr}$
via $z_0 = -a(\tau_0) \sqrt{2\epsilon_{\rm sr}}\Mp$.  For simplicity we describe the scale factor as $a = -1/(H \tau)$ with a constant Hubble rate $H$ during inflation.
 We also indicate with  $\mathcal{H}_0$ the size of the comoving horizon at the time of the transition to the non-attractor era, and with $\Delta N = \ln{(\tau_0/\tau_f)}$ the duration (in e-fold numbers) of the non-attractor phase.} 

\begin{figure}[t!]
\begin{center}
\includegraphics[scale=0.87]{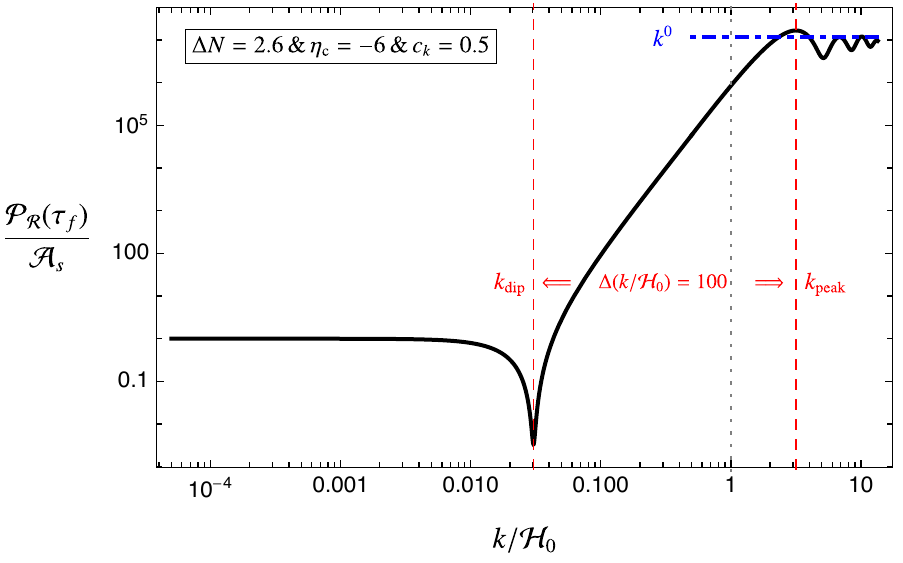}\includegraphics[scale=0.89]{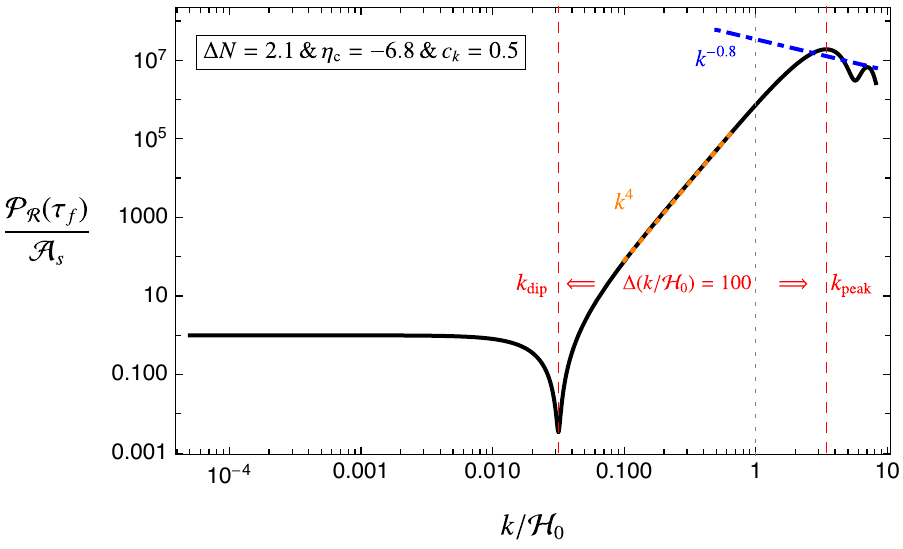}
\end{center}
\caption{Scale dependence of the power spectrum for an example ultra slow-roll model ($\eta_{\rm c} = -6$) (Left) and constant-roll model ($\eta_{\rm c}=-6.8$)(Right) as function of $k/\mathcal{H}_0$.
 The vertical gray-dotted lines separate the interval of wave-numbers leaving the horizon during the initial slow-roll era (left side) and the non-attractor era (right side). We define $\mathcal{A}_s = H^2/(8\pi^2\epsilon_{\rm sr}\Mp^2)$.\label{fig:ps}}
\end{figure}

We  proceed with determining the corresponding growth rate of the power spectrum. For this purpose, we re-write equation  \eqref{psf1} as
\beq\label{psom}
\mathcal{P}_\mathcal{R}(\tau_f,k)  \equiv \left[ (\alpha_k^{R})^2 + (\alpha_k^{I})^2 \right]\mathcal{P}_{\mathcal{R}}(\tau_k)\,	,
\eeq  
and we evaluate the  power spectrum  at $\tau_* \to \tau_f$, \ie at the end of the non-attractor era. Using equation  \eqref{psom}, as well as the expressions \eqref{ar} and \eqref{ai}, we make use of the analytic formulas for $\alpha_k^{R}$ and $\alpha_k^{I}$ from Appendix \ref{AppA} and \ref{AppB} to characterize the shape of the power spectrum. In this way, we plot our results
in  Figure \ref{fig:ps} for  two different sets of parameters characterizing  the instant transition in  the pump field profile of \eqref{zsol1}.
We note that for studying physical implications of our findings, it is convenient to 
 introduce a fixed quantity 
  \beq \label{defOck}
  c_k\,\equiv\,-k\tau_k \, \leq 1\,,
  \eeq
 which  
   determines the size of a mode $k$ with respect to  the horizon $(aH)^{-1}$ at time $\tau = \tau_k$, corresponding to the  horizon crossing epoch. We then
   distinguish modes whose momenta lie in the following ranges:
%
%
%
\begin{itemize}
\item[i)]  modes leaving horizon during the initial slow-roll era, \ie modes satisfying $\tau_k/\tau_0 > 1$ or equivalently $k/\mathcal{H}_0 < c_k \leq 1,$ and
\item[ii)] modes that leave the horizon during the non-attractor $\eta_c \leq -6$ phase, $c_k < k/\mathcal{H}_0 $. 
\end{itemize}

In these regimes, the behaviour of late time $\mathcal{P}_{\mathcal{R}}$ shown in Figure \ref{fig:ps} reflects the accuracy of the gradient expansion formalism in capturing characteristic features of the power spectrum in inflationary backgrounds that transiently violates slow-roll conditions which is required to generate PBHs \cite{Motohashi:2017kbs}. In particular,  we notice a pronounced  dip feature  occurring  at relatively large scales,   associated with modes  that still leave the horizon in the initial slow-roll era (\ie $k_{\rm dip} \ll \mathcal{H}_0$) far away from the peak, which occurs at $k_{\rm peak} \simeq 3\mathcal{H}_0$. 
{It is worth mentioning that such dip feature is due to competing contributions that appear in the power spectrum that are weighted by opposite signs.
In particular, 
 ${\cal P}_\mathcal{R}(\tau_*)$ initially decreases going from large to small scales, because negative terms in eq \eqref{defalk} dominate at small $k$. Positive contributions to late time ${\cal P}_\mathcal{R}$ become instead more and more important at smaller scales (\ie as $k$ increase), where the power spectrum starts to rapidly grow, smoothly connecting with the spectrum in the regime of non-attractor evolution. To sum up, a dip feature in the power spectrum forms around the range of scales where negative and positive contributions to ${\cal P}_{\cal R}$  have comparable magnitude.}
  
Interestingly, the presence of such a pronounced dip in the spectrum is {a universal feature, being} virtually present in all single field models based on non-attractor evolution that are aiming to generate a sizeable peak   in the power spectrum 
 for producing PBH, say of order  $\Delta \mathcal{P}_{\mathcal{R}}/\mathcal{P}_{\mathcal{R}} \simeq 10^{7}$ 
 (see e.g. \cite{Garcia-Bellido:2017mdw,Ezquiaga:2017fvi,Ballesteros:2017fsr,Hertzberg:2017dkh,Cicoli:2018asa,Ozsoy:2018flq,Mahbub:2019uhl,Ballesteros:2020qam,Liu:2020oqe,Kefala:2020xsx}).  
As can be noticed from Figure \ref{fig:ps}, following the scales corresponding to the dip feature, the power spectrum grows with a spectral index that can be as large as $n_s -1 =4$ \cite{Byrnes:2018txb} towards the peak \footnote{Some exceptions to this conclusion can be made through a prolonged non-attractor era $\eta_c = -1$ \cite{Carrilho:2019oqg} (See also \cite{Ozsoy:2019lyy}) which can result with a growth rate of $k^5 (\ln k)^2 > k^4$ or through multiple non-attractor phases that exhibit an almost instantaneous transition \cite{Tasinato:2020vdk}.}. Then,  for modes that leave the horizon deep in the non-attractor era, it decays with a spectral index $n_s -1 = -|6 + \eta_{\rm c}|$\,\,\footnote{Dualities that is present between non-attractor and slow-roll phases \cite{Wands:1998yp,Kinney_2005,Tzirakis:2007bf,Morse:2018kda} can be utilized to extend this behavior to the final slow-roll era during which the inflation must terminate \cite{Atal:2018neu}.}. 

The properties of the dip turn out to be related with global features of the spectrum profile, see e.g. \cite{Tasinato:2020vdk}. In fact, \cite{Tasinato:2020vdk} found that the location of the dip position $ k_{\rm dip}$ in momentum space is related with a characteristic pivot scale $k_\star$ associated with the duration of non-attractor evolution by the expression $ k_{\rm dip}/k_\star\,\simeq\,{\cal O}(1)\times \left( \Delta \mathcal{P}_{\mathcal{R}}/\mathcal{P}_{\mathcal{R}} \right)^{-1/4}$. For the scenarios we consider with a pronounced peak in the power spectrum, we can expect $k_\star$ to be related with $ k_{\rm peak}$, and $\left( \Delta \mathcal{P}_{\mathcal{R}}/\mathcal{P}_{\mathcal{R}} \right)^{-1/4} \simeq  \left( 10^{7} \right)^{-1/4}\,\simeq 10^{-2}$. A close examination of Figure \ref{fig:ps} confirms these arguments and shows a robust relation between the peak scale and the location of the dip feature in momentum space, obeying   
 \beq \label{relpd}
 k_{\rm dip}\,\simeq\,10^{-2}\,{ k_{\rm peak} }
 \,.
 \eeq
 {In light of these considerations, we assume \eqref{relpd}  for the rest of this work. Since the slope of the spectrum changes  abruptly at  the dip position, it is interesting to ask
 whether we can make use of this characteristic feature for designing observables for  probing
 the shape of the spectrum. This is our aim for what comes next.}
 
\begin{figure}[t!]
\begin{center}
\includegraphics[scale=0.89]{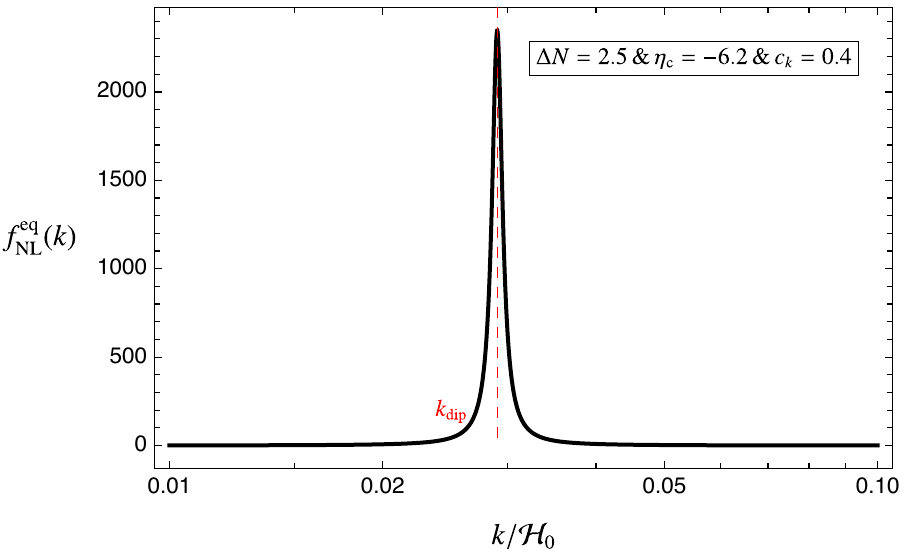}\includegraphics[scale=0.89]{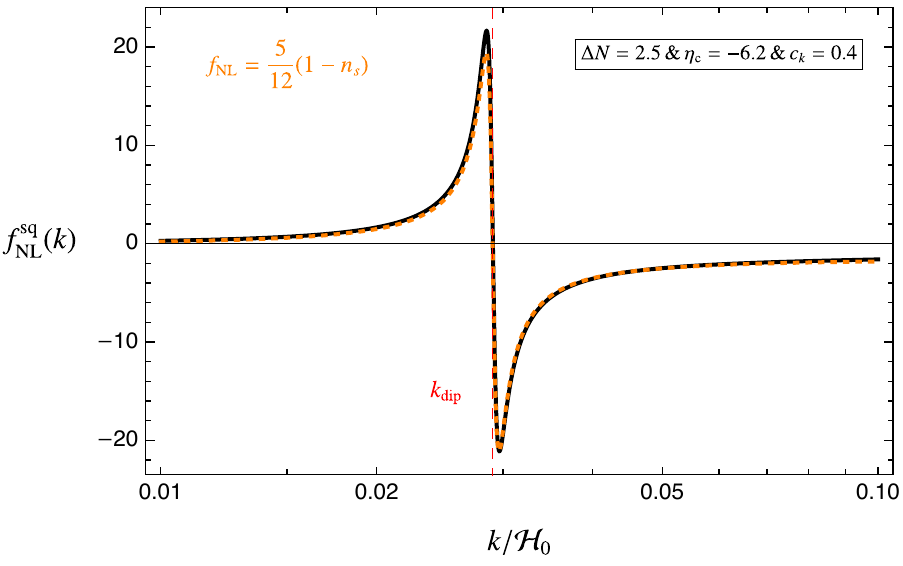}
\end{center}
\caption{Scale dependence of the non-linearity parameter $f_{\rm NL}$ around $k_{\rm dip}$ for the equilateral (Left) and squeezed configuration (Right) for a transient constant-roll model that can generate a $10^7$ enhancement in the power spectrum: $\{\Delta N =2.5,\eta_{\rm c}=-6.2, c_k = 0.4\}$. In the right panel, we represent the accuracy of the consistency relation $f_{\rm NL} = 5(1-n_s)/12$ in capturing the behavior of the $f_{\rm NL}$ in the squeezed limit,  using \eqref{psom} and $n_s - 1\equiv {\d \ln \mathcal{P}_{\mathcal{R}}(\tau_f,k)}/{\d \ln k}$. \label{fig:fnl}}
\end{figure}

\subsection{The scale dependence of the bispectrum}\label{S2p4}

Let us concentrate on modes that exit during the initial slow-roll era, $k/\mathcal{H}_0 < c_k$, to investigate the scale-dependence of the bispectrum. We expect the bispectrum to have features and be amplified around the scale $k_{\rm dip}$ corresponding to the position of the dip in the power spectrum: in fact, non-linearities are usually enhanced at the location of rapid changes in the power spectrum (see e.g. \cite{Chluba:2015bqa} for a review). In what follows, we confirm this expectation for the system under consideration.

For scales satisfying $k/\mathcal{H}_0 < c_k$ (recall the definition of $c_k$ around eq \eqref{defOck}), the power spectrum at around horizon crossing $\mathcal{P}_{\mathcal{R}}(\tau_k)$ is scale invariant, see
 the discussion in Appendix \ref{AppA}, in particular  eq \eqref{pstauk}.  
 We can then simplify the expression for the bispectrum for modes that exit during the initial slow-roll era as (we identify $\mathcal{P}_{\mathcal{R}}(\tau_k) = \mathcal{P}^{(0)}_{\mathcal{R}}$ to emphasize its scale independence)
\beq\label{3PTsr}
B_{\mathcal{R}}\left({k}_{1}, {k}_{2}, {k}_{3}\right)=\frac{(2 \pi^{2})^2}{2(k_{1} k_{2} k_{3})^{3}}\fr{12}{5} f_{\rm NL}(k_1,k_2,k_3) \left[\left|\alpha_{k_1}\alpha_{k_2}\right|^2 k_3^3+\text { perms }\right] (\mathcal{P}^{(0)}_{\mathcal{R}})^2\,.
\eeq
The scale dependent non-linearity parameter~\footnote{A scale-dependent
$f_{\rm NL}$ arises in a variety of other inflationary contexts, see e.g. the early works \cite{Chen:2005fe,Byrnes:2009pe,Byrnes:2010ft}.} then reads as 
\beq\label{fnlsr}
f_{\rm NL}(k_1,k_2,k_3) = \fr{5}{12}\fr{\left[\alpha^{*}_{k_{1}} \alpha_{k_{2}} F(\tau_{k_{3}})\left\{5\left(k_{1}^{2}+k_{2}^{2}\right)-k_{3}^{2}\right\} k_{3}^{3}\,+\text { perms }\right]}{{\left[\left|\alpha_{k_1}\alpha_{k_2}\right|^2 k_3^3+\text { perms }\right]}}.
\eeq
Notice that in eq
 \eqref{3PTsr}, the scale dependence of bispectrum is characterized by the non-linearity parameter $f_{\rm NL}$, and the terms in the square brackets weighted by the enhancement factors $\alpha_k$ contain information on the overall amplification of the power spectrum at late times. In this sense, Eq \eqref{3PTsr} can be interpreted to describe the late time bispectrum in terms of the power spectrum at late times which exhibits an amplification parametrized by $\alpha_k$.

 The equilateral limit of \eqref{3PTsr}  makes these arguments  clear, since  we can re-write  the late time bispectrum  as (here  $ \mathcal{P}_{\mathcal{R}}(\tau_*,k) = \alpha_k \mathcal{P}^{(0)}_{\mathcal{R}}$)
\beq
B_{\mathcal{R}}\left({k}, {k}, {k}\right) = \fr{18}{5} f_{\rm NL}(k,k,k) \,\, \left[\fr{2\pi^2}{k^3} \mathcal{P}_{\mathcal{R}}(\tau_*,k)\right]^2\,.
\eeq
{Nevertheless, we found that other shapes of the bispectrum turn out to be physically interesting. In fact, the bispectrum in eq \eqref{3PTsr} has
a broad support covering different shapes, and our analytical formulas allow us to have analytic control on the various possibilities.  For definiteness, here we discuss the representative equilateral and squeezed configurations, while in Section \ref{S3p3} we will analyze other possible shapes corresponding
to isosceles triangles in momentum space.} 

We now focus on a representative choice of parameters that can generate an enhancement of order $10^{7}$  in the power spectrum -- as required for PBH production. Focusing on such scenarios, we plot in Figure \ref{fig:fnl} the resulting $f_{\rm NL}$ in the equilateral $k_1 = k_2 = k_3$ and squeezed configuration $k_3 \ll k_1 = k_2 $, for scales that exit the horizon during the initial slow-roll era, namely for scales around the $k_{\rm dip}$. From the left panel in Figure \ref{fig:fnl}, we notice that the non-linearity parameter reaches its maximal value at the location of 
 the $k_{\rm dip}$. {The right panel in Figure \ref{fig:fnl} also provides valuable information on the scale dependence of $f_{\rm NL}$ in the squeezed limit. We learn 
 that this quantity initially grows positive, to then  rapidly decrease to negative values around $k_{\rm dip}$, 
 and to finally  increase again from   negative values,   almost in a symmetrical fashion.}  Importantly, as can be seen from the overlap of orange dotted curve with the black solid one in the right panel of Figure \ref{fig:fnl}, the scale dependence of $f_{\rm NL}$ in the squeezed limit agrees very well with the one inferred from Maldacena's consistency condition: $f_{\rm NL} = 5 (1-n_s)/12$.  {The typical  magnitude for the squeezed limit of
 $|f_{\rm NL}|$ around $k_{\rm dip}$ results of order ${\cal O}(10)$, compatible with the large
 value for the spectral tilt around the dip region (see e.g. \cite{Ozsoy:2019lyy}).}
  
 It is worth to point out that the findings we present in Figure \ref{fig:fnl} agree well with  previous literature \cite{Atal:2018neu,Passaglia:2018ixg,Gao:2021vxb} that numerically studied non-Gaussianity in inflationary setups that include a transient non-attractor era. However, the main advantage of our formulas \eqref{psom}, \eqref{3PTsr} and \eqref{fnlsr}  
   is the analytic control they provide us to study features in the power and bispectrum in a model independent way without restoring to numerical techniques. In particular, these formulas are flexible enough to ensure an accurate description of the power and bispectrum in various inflationary backgrounds that exhibit a transient, slow-roll violating phase characterized by essentially two numbers: the duration $\Delta N$ of the non-attractor era and the value of the slow-roll parameter $\eta_c < 0$ in this phase. 

\subsection{Summary of the results so far, and  their  implications}\label{sec-sumim}

Let us briefly summarize our findings so far, and anticipate their implications to be developed
next.  We consider inflationary scenarios
 that incorporate a short non-attractor phase aimed to increase the size of the curvature perturbation spectrum
 at small scales. The spectrum profile is characterized by features, including  a  dip that precedes a steady growth -- see Figure \ref{fig:ps} -- a phenomenon  well studied in previous works.
 Moreover, for the first time we analytically demonstrate that the bispectrum associated with  
  %
 $\mathcal{R}$ acquires  pronounced  features and strong scale dependence around $k_{\rm dip}$ (see Figure \ref{fig:fnl}). 
 
 Can we use
 the pronounced features associated with the statistics of curvature perturbation at
 the scale  $k_{\rm dip}$, occurring     {\bf well away from the peak} (see eq \eqref{relpd}), for probing these 
 inflationary models  with CMB physics only, independently on the details of PBH formation? 
  An affirmative answer 
 would allow us to  probe the PBH generation mechanism within  clean and nearly-linear perturbation theory 
 at relatively large scales, avoiding complications involved with the physics of PBH formation.
 
 \smallskip
 
 With this aim in mind, we  explore the
 idea    to study cross-correlations between  temperature fluctuations at large CMB
 scales, of order $k_L\simeq 10^{-4}-10^{-2} \,{\rm Mpc}^{-1}$, and CMB $\mu$-spectral distortions at smaller scales~\footnote{See \cite{DeLuca:2021hcf} for a recent paper investigating $\mu$-distortions to probe PBH scenarios. The CMB distortions studied in \cite{DeLuca:2021hcf} are caused however by the PBH clustering process after PBH formation, a phenomenon different from the one we consider.}, say $k_S\, \simeq\, 10^{3}-10^{4}\, {\rm Mpc}^{-1}$. Correlations between such disparate scales can indeed be induced by the squeezed limit of the curvature  bispectrum  \cite{Pajer:2012vz}, which in our case is amplified at the dip position to the levels of order $|f_{\rm NL}|\simeq {\cal O}(10)$. Given that for scales $k_{\rm peak} < 10^{5}\,{\rm Mpc^{-1}}$ a peak in the scalar power spectrum is expected to be tightly constrained  from $\langle\mu\rangle$ distortions alone (see e.g. \cite{Byrnes:2018txb,Unal:2020mts}), we focus on  the interval
   $k_{\rm peak} = 10^{5} -10^{6}\, {\rm Mpc}^{-1}$. This interval corresponds to $k_{\rm dip} = 10^{3}-10^{4}\,{\rm Mpc}^{-1}$  (see eq \eqref{relpd}), the range most relevant for $\mu$ distortions,  as we review in the next
   section.    We point out that  selecting  $k_{\rm peak} = 10^{5}-10^{6}\, {\rm Mpc}^{-1}$ leads to the formation
   of  PBHs  in the range $M_{\rm PBH} \simeq 1-100\,M_{\odot} $
    (see e.g. \cite{Garcia-Bellido:2017fdg,Carr:2020xqk}).  PBHs within this mass range are interesting in their own, and moreover 
     can be considered as seeds for Supermassive Black Holes (SMBH) with $M_{\rm SMBH} = 10^{6}-10^{8}\,\,M_{\odot}$, taking into account accretion and merging effects after production \cite{Garcia-Bellido:2017aan,Unal:2020mts}.  However, as one can anticipate from the discussion above, we will not consider any actual details of the PBH production, and in the next section we will concentrate on how a squeezed bispectrum enhances cross-correlations between temperature fluctuationsand $\mu$-type distortions at scales that are much larger than $k_{\rm peak}$.
     
     We conclude this subsection with some comments on the physical significance 
      of the squeezed limit of 
     the bispectrum in single field inflation. There has been some debate in the recent literature on whether 
     such limit is physical, or whether can be  set to zero with a gauge
     transformation. Considering single-field inflationary scenarios in a  non-attractor
     phase,   \cite{Bravo:2020hde} finds that the physical $f_{\rm NL}$  vanishes. On the other
     hand,  
     \cite{Suyama:2021adn} (see also the general discussion in \cite{Matarrese:2020why,Taoso:2021uvl}) 
     points out that the dynamics of the decaying mode can leave important imprints
     in the squeezed limit of the bispectrum, 
     when carefully considering the process of matching between non-attractor and attractor
     regimes towards the end of inflation\footnote{See \eg \cite{Cai:2017bxr} for the implications of this transition on the fate of $f_{\rm NL} \simeq \mathcal{O}(1)$ that can be generated in non-attractor backgrounds such as ultra slow-roll (USR) inflation \cite{Namjoo:2012aa,Martin:2012pe,Chen:2013aj,Chen:2013eea}. In this context, see also \cite{Finelli:2017fml} for a generalized soft theorem in shift symmetric inflationary models including USR inflation where the scalar potential vanishes.  }.
     
     The effects that
     we are going to study hold whenever there is a coupling 
     between long and 
     short modes induced by non-Gaussianity, and does not require  to focus on the 
     infinitely squeezed limit of the bispectrum where the previous  debate  
     applies.  In our case, as we have seen with examples in Section \ref{S2p4}, the bispectrum shape is quite complex and has a broader support than the strictly local shape. Hence it can provide such couplings
     even outside the purely local Ansatz, and we will make use of this fact for 
     our phenomenological considerations in Section \ref{S3p3}.    Moreover, although here we 
     consider single-field systems,  the same phenomenon can also occur in multiple field or curvaton-like scenarios for PBH production
     with pronounced features in the statistics of curvature fluctations. In those cases, the effects of non-adiabatic modes in the squeezed
     limit of the bispectrum 
     can not be removed by gauge transformations. Hence, in what follows we do not discuss
     these issues any further, and  focus on developing the phenomenological consequences of our idea.
     
\section{$\mu T$ correlations as a probe of  the PBH generation mechanism }
\label{sec-pheno}

{ We now discuss  how to use the cross-correlation among CMB spectral distortions and temperature fluctuations for probing the statistics of curvature perturbation in inflationary scenarios capable of generating PBHs. We start with a brief review that discuss $\mu$-type CMB distortions and their properties for testing the statistics of primordial fluctuations following previous works. We then apply these methods to inflationary scenarios leading to PBH production, analyzing the prospect of using $\mu T$ cross correlation for probing the corresponding scale-dependent bispectrum at the dip position. The method we propose allows one to indirectly probe the PBH formation properties at relatively large scales, away from the scales where the curvature spectrum grow and has a peak.} 

\subsection{Brief review of CMB   $\mu$-distortions}

{The energy injection caused by the diffusion damping (Silk damping) of the acoustic waves in the pre-recombination photon-baryon plasma heats photons and leads to spectral distortions in the black-body spectrum of the CMB as they re-enter the horizon \cite{Sunyaev:1970er,Sunyaev:1970er2}. At very high redshifts, these distortions are erased by both photon conserving (\ie Compton scattering) and non-conserving (double Compton scattering) processes. For the range of redshifts $z_f\,\equiv\,5 \times 10^{4}\,<z\,<\, 2 \times 10^{6}\,\equiv\,z_i $, photon  non-conserving processes cease to be efficient but photons can still maintain thermal equilibrium by elastic Compton scattering $e^{-} + \gamma \to e^{-} + \gamma $, which
conserves photon number. The resulting photon spectrum is then described by a Bose-Einstein distribution with a non-vanishing, chemical potential $\tilde{\mu} \equiv \mu/(k_B T)$, leading to the so-called $\mu$-type distortion of the black-body CMB spectrum:
\beq
n(\nu) = [e^{h\nu/(k_BT)}-1]^{-1}\quad \longrightarrow \quad[e^{h\nu/(k_BT)+\tilde{\mu}}-1]^{-1},
\eeq
where $n(\nu)$ is the number density of photons per frequency. Being associated with the dissipation induced in the primordial plasma by the super-horizon primordial fluctuations as they re-enter the horizon and starts oscillating, $\mu$ distortions have primordial origin, and thus can be expressed in terms of the primordial power spectrum 
  \cite{Hu:1994bz,Khatri:2011aj,Chluba:2011hw,Pajer:2012vz}. To see this, we relate the size of $\mu$ distortion to the heat generated by diffusion damping $Q_\gamma$ \cite{Chluba:2012gq}:
\beq
Q_\gamma = \fr{c_s^2}{1+c_s^2} \rho_\gamma \langle \delta_\gamma^2\rangle_p\,\,\,\,,
\eeq
where the square of the sound speed obeys $c_s^2 \simeq 1/3$ since the universe is radiation dominated at those redshifts, and $\langle\dots\rangle_p$ is the square of the photon energy density contrast $\delta_\gamma$ averaged over a period of acoustic oscillations. This energy release is then converted into $\mu$-distortions as
\beq
\mu \simeq 1.4 \int_{z_f}^{z_i} \d z \fr{1}{\rho_\gamma} \fr{\d Q_\gamma}{\d z} \simeq \fr{1.4}{4}  \langle \delta_\gamma^2\rangle_p \bigg|_{z_f}^{z_i}.
\eeq
Denoting the transfer function of photon density contrast $\delta_\gamma (\tau_0,k) = T_\gamma (k) \mathcal{R}_k$ -- with $\mathcal{R}_k$ is the conserved super-horizon curvature perturbation --   $\mu$-distortions are then related to the primordial curvature perturbation by 
\beq\label{mu}
\mu \simeq  \,\,4.6 \int \frac{\d^{3} k_{1} \d^{3} k_{2}}{(2 \pi)^{6}}\,\, \mathcal{R}_{k_1} \mathcal{R}_{k_2} \,e^{i \vec{k}_{+} \cdot \vec{x}}\, W\left(\frac{k_{+}}{k_{s}}\right) \left\langle\cos \left(c_s k_{1} \tau\right) \cos \left(c_s k_{2} \tau\right)\right\rangle_{p} e^{-\left(k_{1}^{2}+k_{2}^{2}\right) / k_{D}^{2}}\bigg|_{z_{f}}^{z_{i}},
\eeq
where  $T_\gamma(k) \simeq 3 \cos(c_s k \tau)\,e^{-k^2/k^2_D}$ \cite{Pajer:2012vz} is the small-scale limit of the linear photon transfer function and $k_D$ denotes the diffusion damping scale. During radiation domination, denoting $R \equiv 3\rho_b/\rho_\gamma \ll 1$ ($\rho_b$ being the baryon density), the damping scale reads as
\beq
k_{D}  \equiv\left[\int_{z}^{\infty} \d z \,\frac{1+z}{6 H n_{e} \sigma_{T}(1+R)}\left(\frac{R^{2}}{1+R}+\frac{16}{15}\right)\right]^{-1 / 2}  \simeq\left(\fr{1+z}{10^5}\right)^{3 / 2} 130\, \mathrm{Mpc}^{-1}.
\eeq

The quantity $k_{D}$ therefore depends on redshift, and will appear in many  formulas  in what follows. 
In \eqref{mu}, $W(k)= 3k^{-3} [\sin(k)-k\cos(k)]$ is top-hat filter function in Fourier space that smears the dissipated energy over a volume of radius $k_s^{-1} \gtrsim k_D(z_f)^{-1}$ and $\vec{k}_\pm = \vec{k}_1 \pm \vec{k}_2$. Using \eqref{mu}, the average (monopole) $\mu-$distortion in the CMB generated by acoustic damping of perturbations in the photon-baryon plasma is then given by the log integral of the primordial power spectrum from $k_D(z_f) \simeq 46\,\, {\rm Mpc^{-1}}$ to $k_D(z_i) \simeq 1.2 \times 10^{4} \,\,{\rm Mpc^{-1}}$: 
\beq\label{muave}
\langle\mu\rangle \simeq 2.3 \int \d \ln k \,\, \mathcal{P}_{\mathcal{R}}(k)\,\,\bigg[e^{-2 k^{2} / k_{D}^{2}}\bigg]_{f}^{i}.
\eeq 
 Eq \eqref{muave}, shows how $\mu-$type distortions\footnote{A slightly different version of the eq. \eqref{muave} is provided in \cite{Chluba:2015bqa,Nakama:2017ohe}:  $\langle\mu\rangle \simeq \int_{k_0}^{\infty} \d \ln k\,\,\mathcal{P}_{\mathcal{R}}(k)\,\, W(k)$ where $W(k) = [\exp(-[\hat{k}/1360]^2/(1+[\hat{k}/260]^{0.3}+\hat{k}/340))-\exp(-[\hat{k}/32]^2)]$ with $\hat{k}\equiv k/{\rm Mpc}$ and $\hat{k}_0 =1$.}, generated by the dissipation of the photon density perturbation at small scales, allow one to probe the primordial power spectrum at small scales. 
  {In particular, $\mu$-distortions can be directly used for constraining inflationary models generating PBHs
  using a pronounced peak in their curvature power spectrum at  scales of order $k_{\rm peak} < 10^{5}\, {\rm Mpc^{-1}}$, see 
  e.g. the recent \cite{Byrnes:2018txb,Unal:2020mts}.
 But as anticipated in Section \ref{sec-sumim}, we can use  spectral distortions to indirectly probe also smaller scales 
 $k_{\rm peak} > 10^{5}\, {\rm Mpc^{-1}}$. Our idea is to  use the large non-Gaussianity produced at the dip position  $k_{\rm dip} \simeq k_{\rm peak}/100$ (see Figure \ref{fig:fnl}), which induces large cross-correlations of $\mu$-distortions with the CMB temperature
 spectrum. As we show in what comes next, the scale-dependence of the bispectrum around the dip feature leads to larger effects compared to the inflationary scenarios that support a purely local bispectrum, making such cross-correlations a valuable observable for testing mechanisms of PBH generation independently from the details of the PBH production.
 }

\subsection{Cross-correlation of $\mu$-distortions with CMB temperature anisotropies}

We now briefly review how primordial non-Gaussianity induces cross-correlation between $\mu$-distortions and CMB temperature anisotropies, closely following \cite{Pajer:2012vz,Ganc:2012ae}. We then discuss and compare the resulting $\langle \mu T \rangle$ correlation focusing on two different cases: i) for a primordial scenario that can generate a purely local type bispectrum and ii) for a slow-roll violating inflationary model that features a scale dependent squeezed bispectrum around the dip feature as we discussed in Sections \ref{S2p3} and \ref{S2p4}.
   
As first shown in \cite{Pajer:2012vz}, a non-zero bispectrum in the squeezed limit makes the distribution of $\mu$ anisotropic on the sky and allows cross-correlations between $\mu$ anisotropy and CMB temperature anisotropy $\Theta(\hat{n}) = \delta T(\hat{n})/\bar{T}$. To see this, we can expand the CMB temperature anisotropies detected by an observer into spherical harmonics as $\Theta(\hat{n}) = \sum_{lm} a^{T}_{lm}\,Y_{lm}(\hat{n})$, where the harmonic coefficients are given by
\beq\label{aT}
a_{l m}^{T} \equiv \int \d \hat{n}\, \Theta(\hat{n}) \,Y_{l m}^{*}(\hat{n}) = \frac{12 \pi}{5}(-i)^{l} \int \frac{\d^{3} k}{(2 \pi)^{3}} \mathcal{R}_k\, \Delta_{l}(k)\, Y_{l m}^{*}(\hat{k}),
\eeq
and $\Delta_l$ is the radiation transfer function. Similarly, for an observer at the origin, direction dependent distortion anisotropies in $\mu(\hat{n})$ can be expanded into spherical harmonics as $\mu(\hat{n}) = \sum_{lm} a^{\mu}_{lm}\,Y_{lm}(\hat{n})$. Using eq. \eqref{mu}, we can then relate the the spherical harmonic coefficients to the primordial curvature perturbation as \cite{Pajer:2012vz,Ganc:2012ae}
\begin{align}\label{amu}
\nn a^{\mu}_{lm} &\simeq18.4 \pi(-i)^{l} \int \frac{\d^{3} k_{1} \d^{3} k_{3}}{(2 \pi)^{6}}\,\, Y_{l m}^{*}(\hat{k}_{+}) \,\,\mathcal{R}_{k_1} \mathcal{R}_{k_2} \,\,W\left(\frac{k_{+}}{k_{s}}\right) j_{l}(k_{+} \chi_{*}) \\
&\quad\quad\quad\quad\quad\quad\quad\quad\quad\quad\quad\quad\times \langle\cos \left(c_s k_{1} \tau \right)\cos \left(c_s k_{2} \tau\right)\rangle_{p}\left[e^{-\left(k_{1}^{2}+k_{2}^{2}\right) / k_{D}^{2}}\right]_{f}^{i}\,,
\end{align}
where $\chi_* = \tau_0-\tau_*\simeq 14\, {\rm Gpc}$ is the comoving distance between the last scattering surface and today. Defining the angular correlators of two anisotropies labeled by $\{i,j\}$ as
\beq
\left\langle\left(a_{l m}^{i}\right)^{*} a_{l^{\prime} m^{\prime}}^{j}\right\rangle=\delta_{l l^{\prime}} \delta_{m m^{\prime}} C_{l}^{i j}.
\eeq 
The cross correlation of \eqref{amu} and \eqref{aT} then gives
\beq\label{CLmuT}
C_{l}^{\mu T}\simeq \frac{27.6}{160 \pi^{3}} \int \d k_{+}\,k_{+}^{2}\, \Delta_{l}(k_{+})\,j_{l}\left(k_{+} \chi_*\right)W\left(\frac{k_+}{k_{s}}\right)\int  \d k_{-}\,k_{-}^{2}\,  B_{\mathcal{R}}\left(\fr{k_{-}}{2}, \fr{k_{-}}{2}, k_+\right)  \left[e^{-(k_{-}^{2}+k_{+}^2) /2 k_{D}^{2}}\right]_{f}^{i}\,.
\eeq
In this expression 
 we  choose the smoothing scale to be the same as the largest scale $\mu$ distortions are relevant for, \ie $k_s\approx k_D(z_f)$ and focused on the squeezed limit $k_{+} = k_3 \to 0$, $k_1 \to k_{-}/2$ together with the definition of the bispectrum in \eqref{BSdef}. It is clear from \eqref{CLmuT} that the angular correlator $C^{\mu T}_l$ is sensitive to the integral of the  bispectrum in the squeezed configuration, in particular with a larger ratio $k_1/k_3 \simeq k_{-}/2k_{+}$ compared to the CMB anisotropies alone. Noting that the smallest wave-number we can probe from the CMB anisotropy in the sky corresponds to the quadrupole, $k_3 \approx l/ \chi_* = 2/(14\,{\rm Gpc})\simeq 1.4 \times 10^{-4}\,{\rm Mpc^{-1}}$, we have  access to the ratio $k_1/k_3$ within the following range,
\beq
\fr{k_D(z_f)}{k_3} < \fr{k_1}{k_3} < \fr{k_D(z_i)}{k_3}\quad\quad \Rightarrow \quad\quad 3.2\times 10^{5}<\fr{k_1}{k_3} <8.4\times 10^{7}.
\eeq
Note that this ratio is much greater one can reach only via temperature anisotropies of the CMB for $l = 2-3000$ where $k_1/k_3 = l_1/l_3 = 1-1500$. To set the stage for the computation of $\langle \mu T \rangle$ in inflationary scenarios that can produce large PBH populations in the post-inflationary universe (See Sections \ref{S2p3} and \ref{S2p4}), below we first review the standard calculation for a purely local bispectrum.
\subsection{$C^{\mu T}_{l}$ for a local type bispectrum}\label{xcorsr}
For local type non-Gaussianity, the bispectrum of the curvature perturbation in the squeezed limit can be expressed as \cite{Komatsu:2001rj}~\footnote{See Appendix \ref{AppC} for a derivation on the single field slow-roll consistency relation $f^{(p)}_{\rm NL} = 5 (1-n_s)/12$ within the framework of super-horizon gradient formalism we are focusing in this work.}   
\beq\label{Bsr}
B^{\rm loc}_{\mathcal{R}}\left(\fr{k_{-}}{2}, \fr{k_{-}}{2}, k_+\right) = \fr{12}{5} f^{(\rm p)}_{\rm NL}\,\, \left[\fr{2\pi^2}{(k_{-}/2)^{3}} \mathcal{P}_{\mathcal{R}}(\tau_{k_{-}/2})\right] \left[\fr{2\pi^2}{k_{+}^{3}} \mathcal{P}_{\mathcal{R}}(\tau_{k_{+}})\right],
\eeq
where   $f^{(\rm p)}_{\rm NL}$ is the scale-independent primordial non-linearity parameter \footnote{See \cite{Biagetti:2013sr,Chluba:2016aln} for earlier works on the study of cross correlations between $\mu$ and $\Theta$ in the presence of scale dependent $f_{\rm NL}$.}. Adopting for simplicity a scale invariant primordial power spectrum  $\mathcal{P}^{(0)}_{\mathcal{R}} = 2.1 \times 10^{-9}$, and plugging \eqref{Bsr} into \eqref{CLmuT}, we obtain\footnote{It should be noted that there are non-primordial contributions to the final observed $C_l^{\mu T}$ from non-linear and projection effects which are found to be negligible \cite{Cabass:2018jgj}.}
\beq\label{CLmuTsr}
C_{l}^{\mu T} \simeq  2.7 \times 10^{-17}\, f^{(\rm p)}_{\rm NL}\, \,\fr{2\pi}{l(l+1)}b_{(\rm loc)}(l)\,,
\eeq
where the multipole dependent quantity  ${b_{(\rm loc)}}$ (the suffix $(\rm loc)$ means that it holds for a local bispectrum) can be expressed in terms of a double integration over short and long momenta involving the transfer function $\Delta_l$ of temperature anisotropies and window function $W$ as%
\beq \label{CLmuTsra}
{b_{(\rm loc)}}(l)\,=\,
  \fr{6\, l (l+1)}{\ln\left(\fr{k_D (z_i)}{k_D(z_f)}\right)} \int \d \ln k_{+}\, \Delta_{l}(k_{+})\,j_{l}\left(k_{+} \chi_*\right)\, W\left(\frac{k_+}{k_{s}}\right) \int  \d \ln k_{-}\,  \left[e^{-{(k_{-}^{2}+k_{+}^2)}/{2 k_{D}^{2}}}\right]_{f}^{i}.
\eeq
To compute this quantity, one can first focus on the Sachs-Wolfe (SW) limit (small $l$) where $\Delta_l \to j_l(k_+\chi_*)/3$. Then realizing that $k_+ \ll k_s \simeq k_D(z_f)$ and $W(k_+/k_s) \to 1$ in the squeezed limit $k_+ \to 0$, the integrals in \eqref{CLmuTsra} can be carried analytically to conclude $b_{(\rm loc)}\to1$ and arrive at the standard result
\beq\label{CLmuTsrf}
C_{l}^{\mu T, {\rm SW}} \simeq 2.7 \times 10^{-17}\, f^{(\rm p)}_{\rm NL}\,\fr{2\pi}{l(l+1)},
\eeq
\begin{figure}[t!]
\begin{center}
\includegraphics[scale=0.99]{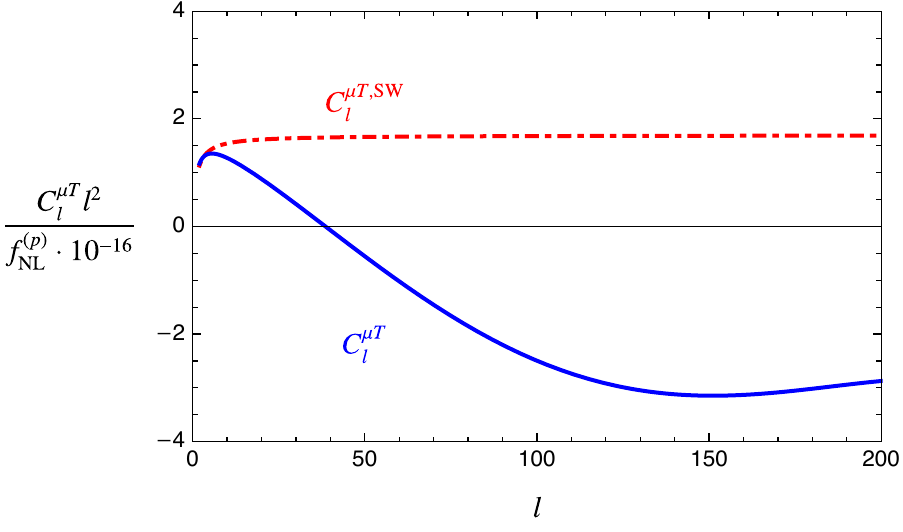} 
\end{center}
\caption{Exact scale dependence of $C^{\mu T}_l$ for a local type bispectrum \eqref{Bsr} including full transfer function effects, see \eg \eqref{rhol} (solid line). $C^{\mu T}_l$ that arise in the Sachs-Wolfe limit (see \eg \eqref{CLmuTsrf}) is shown by dotted dashed line.\label{fig:clsr}}
\end{figure}\noindent
which indicates that on large scales (where the SW approximation is valid) $C_{l}^{\mu T}$ is scale invariant as $l(l+1) C_{l}^{\mu T}= {\rm constant}$.

To obtain the exact $l$ dependence of $C_{l}^{\mu T}$ in \eqref{CLmuTsr}, one needs to include the full transfer function $\Delta_l(k_+)$ of temperature anisotropies in the integral  \eqref{CLmuTsra} over long momenta $k_+$, which can be implemented numerically \cite{Ganc:2012ae}. It turns out that the final $C^{\mu T}_l$ obtained in this way can be related to $C^{\mu T,SW}_l$ by an analytic fit first proposed in \cite{Chluba:2016aln}:
\begin{eqnarray}
\label{rhol}
C^{\mu T}_{l} &=& \rho(l)\,\, C^{\mu T, {\rm SW}}_l\,, \hskip1cm {\rm with}
\\
\rho(l) &\simeq& 1.08\left[1- 0.022\,l-1.72\times 10^{-4}\,l^2+2\times 10^{-6}\,l^3-4.56 \times 10^{-9}\,l^4\right].
\nonumber
\end{eqnarray}
The resulting $C^{\mu T}_{l}$ in comparison with $C^{\mu T, {\rm SW}}_{l}$  is shown in Figure \ref{fig:clsr}. As first shown in \cite{Ganc:2012ae}, the SW approximation breaks down around $l =10$ and $C^{\mu T}_l$ changes sign at $l \simeq 40$ becomes negative for $l > 40$. This implies that for a local type bispectrum with $f^{(\rm p)}_{\rm NL} >0$, while temperature anisotropies and $\mu$ distortions are correlated at the largest scales $l < 40$, they become anti-correlated at small scales. 
\subsection{$C^{\mu T}_{l}$ for inflationary scenarios with an enhanced power spectrum} \label{S3p3}
We next focus on inflationary models that contains a transient non-attractor era $\eta_c < 0$ following the initial slow-roll phase, as discussed in Section \ref{SecHeu}.
 Using the gradient expansion method of Section \ref{SecHeu}, eqs
   \eqref{3PTsr} and \eqref{fnlsr}, for modes that exit the horizon in the initial slow-roll era, we obtain the following analytic expression for the  bispectrum associated
with  isosceles triangle configurations:
\beq\label{BSna}
B_{\mathcal{R}}\left(q, q, {k}\right)=\left\{|\alpha_{q}|^2 F(\tau_{k})k^2\left[5-\fr{1}{2}\fr{k^2}{q^2}\right]\fr{k}{q}\,+\alpha_q^* \alpha_{k} F(\tau_q) q^2\left[4 + 5 \fr{k^2}{q^2}\right]  \right\} P_\mathcal{R}(q) P_\mathcal{R}(k),
\eeq
where $P_\mathcal{R}(k)= 2\pi^2 \mathcal{P}^{(0)}_{\mathcal{R}}/ k^3$ is the dimensionful power spectrum at around horizon crossing $\tau_k$ during the initial slow-roll era. Notice that -- as opposed to the parametrization we undertake in eq \eqref{deffnl} --  in \eqref{BSna} we express the bispectrum with respect to two copies of the power spectrum, evaluated at horizon crossing. This way of expressing  the bispectrum provides an easier comparison with the local bispectrum case we reviewed earlier, since eq. \eqref{BSna} can be recast in a form that resembles the standard local type bispectrum for isoceles triangle configurations:
 \beq\label{bsna}
B_{\mathcal{R}}(q,q,k) = \fr{12}{5} f^{\rm eff}_{\rm NL}(q,q,k) P_{\mathcal{R}}(q)\, P_{\mathcal{R}}(k),
\eeq
where the effective non-linearity parameter is given by
\beq\label{fnleffna}
f^{\rm eff}_{\rm NL}(q,q,k) = \fr{5}{12} \left\{|\alpha_{q}|^2 F(\tau_{k})k^2\left[5-\fr{1}{2}\fr{k^2}{q^2}\right]\fr{k}{q}\,+\alpha_q^* \alpha_{k} F(\tau_q) q^2\left[4 + 5 \fr{k^2}{q^2}\right]  \right\}.
\eeq
In appendix \ref{AppD}, we study the squeezed limit $k \ll q$ of \eqref{fnleffna} and found that it quickly reaches to a small constant value $f^{(0)}_{\rm NL}$ (see \eg \eqref{fnleffappf}) in the large scale tail of the squeezed limit, $q\to 0$ and $q > k$. Since, we send $q\to 0$ to reach this initial value, we do not expect that $f^{(0)}_{\rm NL}$ can reflect an accurate initial value of the squeezed configurations of the expression \eqref{fnleffna}. To keep our discussion on the observability of $ \langle \mu T \rangle$ (see Section \ref{S3p5}) as general as possible, we will therefore normalize $f^{\rm eff}_{\rm NL}$ with respect to this initial value (as in \eqref{fnleffappf}) and rescale it with a fiducial primordial $f^{(\rm p)}_{\rm NL}$ that parametrizes the effective local non-linearity parameter valid on large scales \ie at CMB scales. Following appendix \ref{AppD}, in the squeezed limit $k\ll q$, we will therefore adopt the following $f^{\rm eff}_{\rm NL}$
\begin{figure}[t!]
\begin{center}
\includegraphics[scale=0.88]{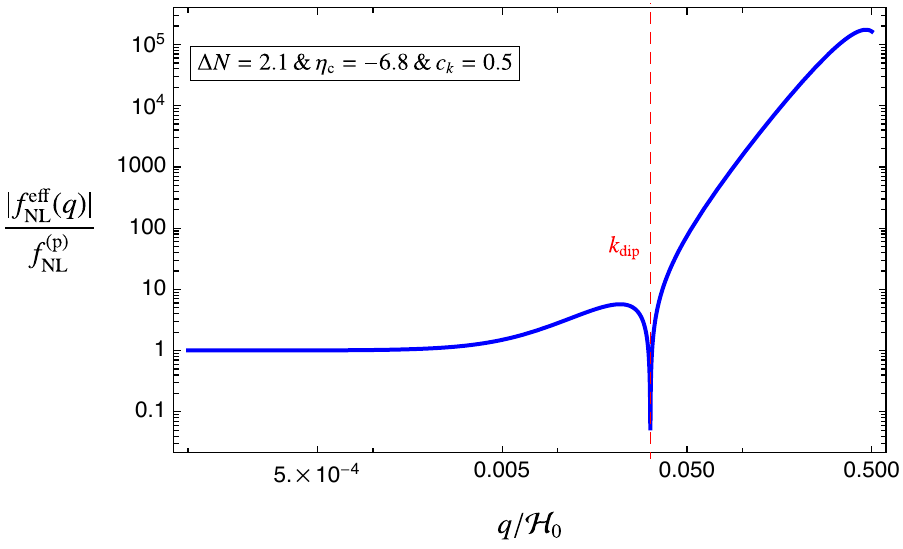} \includegraphics[scale=0.85]{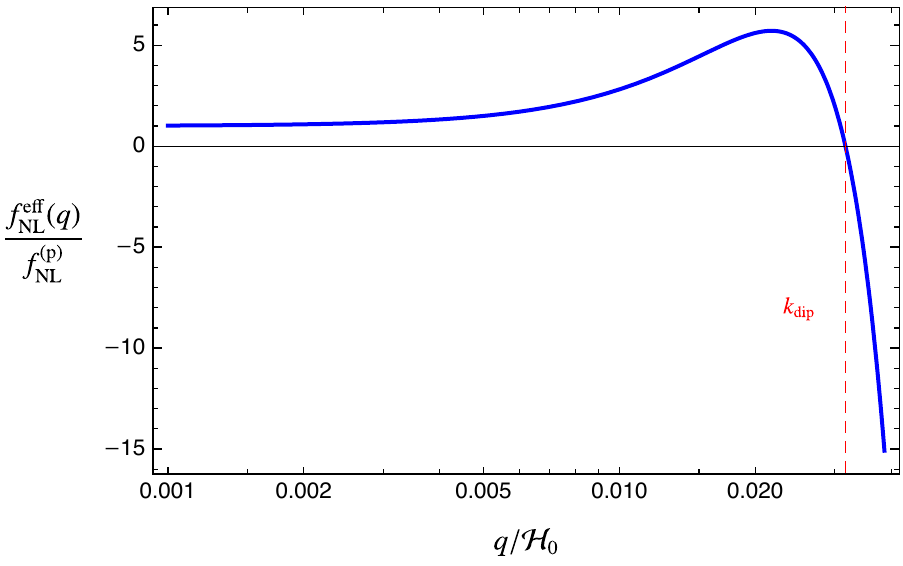}
\end{center}
\caption{Scale dependence of the effective non-linearity parameter $f^{\rm eff}_{\rm NL}$ \eqref{fnleffna} in the squeezed limit $k_3/q \to 0$ for an inflationary model that exhibit an intermediate constant-roll phase $\eta_{\rm c}=-6.8$ that lasts $\Delta N = 2.1$ e-folds.\label{fig:fnleff}}
\end{figure}
\beq\label{fnleffF}
f^{\rm eff}_{\rm NL}(q,q,k) \equiv f^{(\rm p)}_{\rm NL}\, \bar{f}^{\rm eff}_{\rm NL} (q,q,k) \simeq f^{(\rm p)}_{\rm NL}\left[ 1 +\sum^{6}_{n=2} \fr{c^{(n)}_{{\rm NL}}(\eta_{\rm c},\Delta N,c_q)}{c^{(0)}_{{\rm NL}}(c_q)}\left(\fr{q}{\mathcal{H}_0}\right)^{n}\right] + \mathcal{O}\left(\fr{k}{q}\right),
\eeq
where $\mathcal{H}_0 \simeq k_{\rm peak}/3 \simeq 100 k_{\rm dip} /3$ (see Section \ref{S2p3}) and the size of the coefficients $c^{(n)}_{\rm NL}$ depends on the underlying model
of inflation leading to PBH production characterized by the parameters such as $\{\eta_c, \Delta N\}$. For the precise dependence of these coefficients on the model parameters and quantities relevant to the gradient expansion formalism, see eqs. \eqref{d3}-\eqref{d8}.

Notice that the parametrization \eqref{fnleffF} ensures $f^{\rm eff}_{\rm NL} \to f^{({\rm p})}_{\rm NL}$ in the large scale limit $q \to 0$. On the other hand,  when we approach to smaller scales with $q\,\simeq \,k_{\rm dip}$, the scale dependence of the bispectrum
 becomes important, and its features relevant around the $k_{\rm dip}$. To illustrate this, we present the scale dependence of $f^{\rm eff}_{\rm NL}(q,q,k)$ \eqref{fnleffF} in the squeezed limit ($k/q \to 0$) in Figure \ref{fig:fnleff}. As expected, $f^{\rm eff}_{\rm NL} \to f^{({\rm p})}_{\rm NL}$ in the large-scale tail of the squeezed limit ($q \to 0$ where $q > k_3$). More importantly, we observe that $f^{\rm eff}_{\rm NL}$ changes sign at $k_{\rm dip}$ and starts to grow large in absolute value in the negative direction.
We now turn our attention to the $\langle \mu T \rangle$ (\eqref{CLmuT}) cross correlation that arise through the scale dependence of such bispectrum around the scales $k_{\rm dip}$ associated with the dip feature. For this purpose, noting the definition in \eqref{fnleffF}, we utilize \eqref{bsna} in \eqref{CLmuT}. We then obtain the angular $\mu T$ cross correlation as
\beq\label{CLmuTnaf}
C_{l}^{\mu T}  \simeq 2.7 \times 10^{-17}\, f^{(\rm p)}_{\rm NL}\,\fr{2\pi}{l(l+1)} b_{(\rm pbh)}(l) \,,
\eeq
where we defined  the quantity $b_{(\rm pbh)}$, whose expression in the inflationary scenarios containing a non-attractor phase can be re-written as
\beq\label{b}
{b_{(\rm pbh)}}(l) \equiv \fr{6\, l (l+1)}{\ln\left(\fr{k_D (z_i)}{k_D(z_f)}\right)}\int \d \ln k \, \Delta_{l}(k)\,j_{l}\left(k \chi_*\right)\, W\left(\frac{k}{k_{s}}\right)\int  \d \ln q\, \bar{f}^{\rm eff}_{\rm NL}\left(q,q, k \right) \left[e^{-2q^2/{k_{D}^{2}(z)}}\right]_{f}^{i}.
\eeq

We note that the expression above only assumes a bispectrum in the isosceles triangle configurations without specifying a specific triangle shape such as equilateral or squeezed configurations. This quantity ${b_{(\rm pbh)}}$ involves a double integral over short and long momenta, and coincides with 
expression $ {b_{(\rm loc)}} $ of equation \eqref{CLmuTsra} if the effective non-linearity parameter does not exhibit scale dependence, \ie $\bar{f}^{\rm eff}_{\rm NL} \to 1$. However, as we mentioned previously, for an inflationary scenario that can generate PBH populations, the effective $f^{\rm eff}_{\rm NL}\left(q,q, k\right)$ \eqref{fnleffF} is scale-dependent especially around the dip feature present in the power spectrum (See \eg Figure \ref{fig:fnleff}). This situation might lead to substantial differences between $ {b_{(\rm pbh)}}$  and  $ {b_{(\rm loc)}}$ that can allow us to probe the underlying PBH production scenario.
  
 In particular, we expect that these differences can appear for two reasons:
  \begin{itemize}
  \item[i)] The (possible) dependence of  $\bar{f}^{\rm eff}_{\rm NL}\left(q, q, k\right)$ on the small momentum
  $k$, corresponding to CMB temperature anisotropy scales. Since $k$  also appears in the $l$-dependent functions $ \Delta_{l}$, $j_l$ in the first
  of the nested integrals in \eqref{b}, such a dependence can potentially change the multipole dependence of $b_{(\rm pbh)}$ and hence the resulting $C^{\mu T}_l$ correlator.
  \item[ii)] The dependence of  $\bar{f}^{\rm eff}_{\rm NL}$ on the momentum $q$, corresponding to scales probed by spectral distortions. Such a dependence would only modify the second of the nested integrals in eq  \eqref{b}, changing the overall magnitude of $b_{(\rm pbh)}$ but not its $l$-dependence.
  \end{itemize}
  For what respects point i) above, in this work we focus on the bispectrum in the squeezed limit. Focusing on such configurations, we found that at leading order in the small parameter $k/q \ll 1$,  $\bar{f}^{\rm eff}_{\rm NL}$ is independent from the soft momenta $k$ as can be verified from \eqref{fnleffF} (See also appendix \ref{AppD}). This implies that, in the ultra-squeezed limit, the $l$-dependence of the resulting $C^{\mu T}_l$ is not modified compared to the local bipsectrum case we studied earlier (see Section \ref{S3p3}). Instead, for what respects to point ii), we found that the dependence of  $\bar{f}^{\rm eff}_{\rm NL}$ on the large momenta $q$ is quite strong (See eq. \eqref{fnleffF}) such that it can enhance the overall magnitude of $ {b_{(\rm pbh)}} $,  improving the chances of testing the effects of non-attractor periods in models with PBH production. 
  
Based on these facts, we expect that the $l$-dependence of eq. \eqref{b} will be the same as in the local bispectrum case. Therefore, the full multipole ($l$) dependence of $b_{(\rm pbh)}$ (or $C^{\mu T}_l$ \eqref{CLmuTnaf}) can be obtained starting from a computation made in the large-scale SW limit which can then be extended to smaller scales by means of the analytic fit of eq \eqref{rhol} using $b_{(\rm pbh)}(l) = \rho(l)\, b^{(\rm SW)}_{(\rm pbh)}$. For this purpose, we first focus on the expression $b_{(\rm pbh)}$ of \eqref{b} in the SW approximation,  assuming $\Delta_l(k) \to j_l(k \chi_*)/3$ in this limit. Using the leading order result \eqref{fnleffappf} for $\bar{f}^{\rm eff}_{\rm NL}$ in the squeezed limit, the integrals in \eqref{b} thus give 
\beq\label{bf}
b^{(\rm SW)}_{(\rm pbh)} = 1 + \fr{1}{\ln\left(\fr{k_D (z_i)}{k_D(z_f)}\right)} \sum^{6}_{n=2} \fr{c^{(n)}_{{\rm NL}}}{c^{(0)}_{{\rm NL}}}\fr{\Gamma(n/2)}{2^{(n+2)/2}}\left(\fr{k_D(z)}{\mathcal{H}_0}\right)^{n}\Bigg|^{z_i}_{z_f},
\eeq

\begin{figure}[t!]
\begin{center}
\includegraphics[scale=0.88]{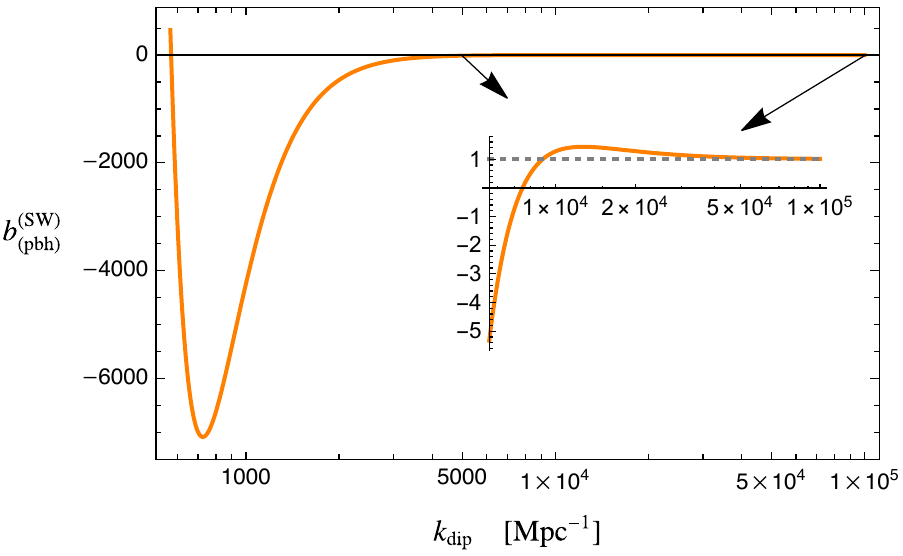}\includegraphics[scale=0.88]{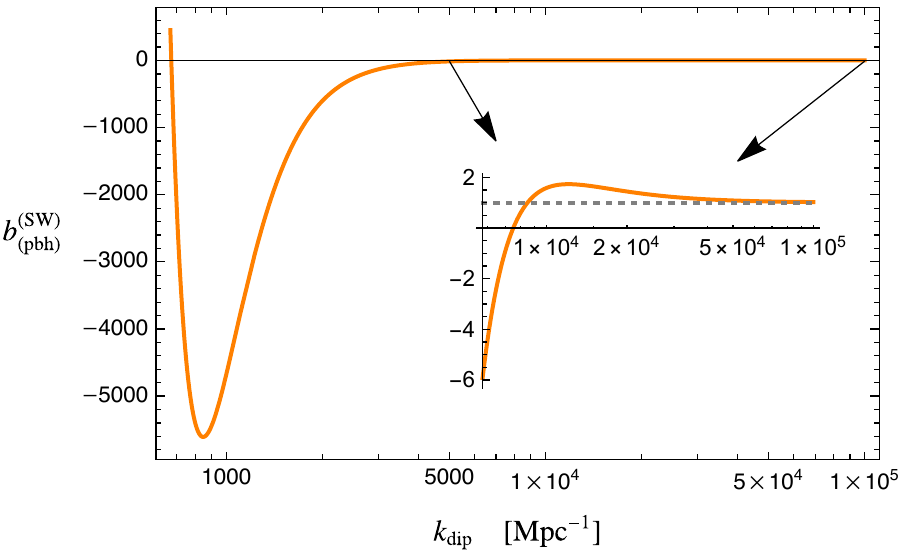} 
\end{center}
\caption{Parameter $b^{(\rm SW)}_{(\rm na)}$ \eqref{bf} as a function of the location of the dip feature $k_{\rm dip} \simeq 3 \mathcal{H}_0/100$ in an inflationary model that contains a transient non-attractor phase parametrized by the parameter set $\{ \Delta N = 2.6, \eta_{\rm c}=-6, c_k = 0.5\}$ (Left) and $\{ \Delta N = 2.5, \eta_{\rm c}=-6.2, c_k = 0.4\}$ (Right).\label{fig:b}}
\end{figure}
\begin{figure}[t!]
\begin{center}
\includegraphics[scale=0.98]{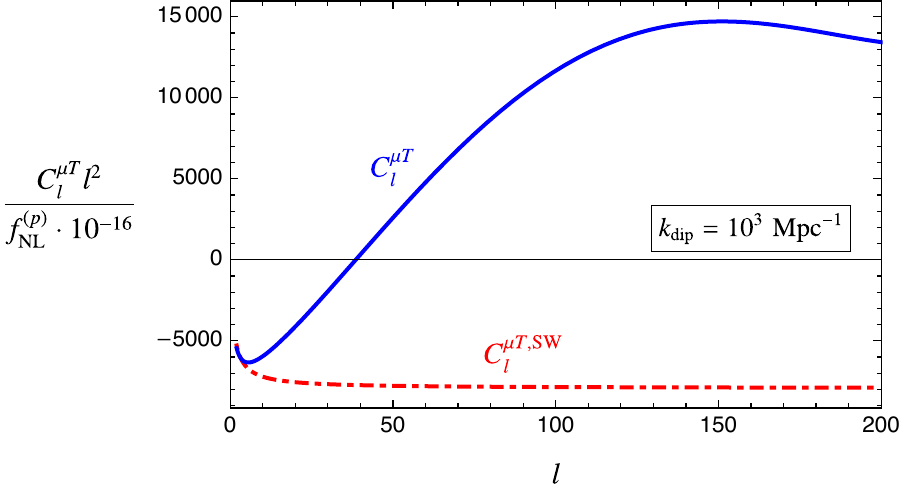} 
\end{center}
\caption{$C^{\mu T}_l$ \eqref{CLmuTnaf} in an inflationary model that contains a transient constant-roll era ($\eta_{\rm c}=-6.2$) that has a duration of $\Delta N = 2.5$ e-folds where we assume $k_{\rm dip} = 10^3\,{\rm Mpc}^{-1}$.\label{fig:clna}}
\end{figure}\noindent
where the dependence of the coefficients on the background parameters in the non-attractor era, \ie $\{\eta_{\rm c}, \Delta N\}$ can be obtained from the formulas we provide in Appendix \ref{AppD}, \ref{AppB} and \ref{AppA}. Armed with \eqref{bf}, we can quantify the extent of which the scale dependence in the bispectrum can influence the $\mu T$ angular correlator $C^{\mu T}_l$ \footnote{It is worth emphasizing again that the leading order contribution we found in \eqref{bf} can only influence the overall amplitude of \eqref{CLmuTnaf} without introducing any scale dependence (\ie $l$ dependence) to the angular cross correlator. This is clear from \eqref{bf} and, as we mentioned above it stems from the fact that at leading order in the squeezed limit $k/q \to 0$, $\bar{f}^{\rm eff}_{\rm NL}$ in \eqref{b} does not exhibit a dependence on the soft momenta $k$ (See Appendix \ref{AppD}).}. For this purpose, we use \eqref{bf} to compute $b^{(\rm SW)}_{(\rm pbh)}$ in terms of the location of the dip in momentum space $k_{\rm dip}$ by noting the relationship $\mathcal{H}_0 \simeq 100 k_{\rm dip} /3 \simeq 33\, k_{\rm dip}$ we derived in Section \ref{S2p3}. The resulting dependence of $b^{(\rm SW)}_{(\rm pbh)}$ on the location of the dip feature $k_{\rm dip}$ in shown in Figure \ref{fig:b}
for two different choices of parameter set that describes an inflationary model including a transient non-attractor era. We observe that for phenomenologically interesting values of $k_{\rm dip}$, $b^{(\rm SW)}_{(\rm pbh)}$ can reach to negatively large values and hence influence the \emph{overall amplitude} of the $\langle \mu T \rangle$ correlator significantly, in particular compared to a inflationary scenario that can generate a scale independent local type bispectrum where $b^{(\rm SW)} \to 1$ as in \eqref{CLmuTsrf}.
 
To illustrate this amplification and the full $l$ dependence of $\langle \mu T \rangle$, in Figure \ref{fig:clna} we plot $C^{\mu T}_l$ \eqref{CLmuTnaf} for an inflationary model that contains an intermediate non-attractor phase with order one negative $\eta_{\rm c}$, both within the Sachs-Wolfe approximation (dot-dashed line), and by taking into account full transfer effects (solid line) using $b_{(\rm pbh)}(l) = \rho(l)\, b^{(\rm SW)}_{(\rm pbh)}$ with \eqref{rhol}. We observe that compared to the scale independent local bispectrum case (See Figure \ref{fig:clsr}), the overall amplitude of the angular correlator is enhanced significantly and its behavior from large to small scales is inverted due to the negativity $b^{(\rm SW)}_{(\rm pbh)}$ for the $k_{\rm dip}$ value quoted in Figure \ref{fig:clna}. In other words, for a PBH forming inflationary scenario, $\mu$ distortions become anti-correlated with temperature anisotropies at large scales contrary to the case arise for a scale independent local bispectrum. It is worth stressing that this behavior arise because in PBH forming inflationary scenarios, scale dependent bispectrum changes sign around the dip feature and grows large in the negative direction as we show in the right panel of Figure \ref{fig:fnleff}. This result is interesting on its own as one can in principle distinguish between these scenarios by just looking at the sign of the $C^{\mu T}_l$ at large scales, \ie for small $l$. In summary, our results imply that phases
of non-attractor inflation can qualitatively and quantitatively change the properties of the cross-correlations between CMB temperature anisotropies and $\mu$-type distortions.

\subsection{Prospects of detectability for $\langle \mu T \rangle$}\label{S3p5}
In order to asses the prospects of detectability  for the $\mu T$ correlator, we  estimate the cumulative signal-to-noise ratio, $S/N$ using \cite{Ganc:2012ae}
\beq\label{StoNdef}
\left(\frac{S}{N}\right)^{2}=\sum_{l = 2}^{l_{\max }}(2 l+1) \frac{\left(C_{l}^{\mu T}\right)^{2}}{C_{l}^{T T} C_{l}^{\mu \mu, N}}\,\,,
\eeq
where $C^{TT}_l$ is the CMB temperature anisotropy power spectrum and $C^{\mu\mu, N}_l$ is the noise level for $\mu$ distortions. For an experiment as PIXIE \cite{Kogut:2011xw}  one has  $C^{\mu\mu, N}_l \simeq 4\pi \, \mu_{\rm min}^2\, e^{l^2/84^2}$ \cite{Pajer:2012vz} where $\mu_{\rm min}$ denotes the minimum detectable $\mu$ distortion (monopole) signal. Using \eqref{aT}, we  denote the $TT$ correlator in \eqref{StoNdef} as
\beq\label{CLTT}
C_l^{TT} = \fr{36\pi}{25} \int \d \ln k \, \mathcal{P}_{\mathcal{R}}(k) \Delta_l^2(k).
\eeq
Following our discussion in the previous section, taking into account full transfer effects we express the $\mu T$ cross correlation as 
\beq\label{cl}
C^{\mu T}_l \simeq 2.7 \times 10^{-17} f^{(\rm p)}_{\rm NL} \fr{2\pi}{l(l+1)} b^{(\rm SW)}_{(\rm pbh)} \rho(l)\,,
\eeq
with $\rho(l)$  defined in \eqref{rhol}.  Then we plug \eqref{cl} into \eqref{StoNdef} to re-write $(S/N)^2$ ratio as
\beq\label{StoN2}
\left(\fr{S}{N}\right)^2 \simeq 7.3\pi \times 10^{-18}  \left[f^{(\rm p)}_{\rm NL}  b^{(\rm SW)}_{(\rm pbh)}\right]^2 \left(\fr{10^{-8}}{\mu_{\rm min}}\right)^2\,\, \sum_{l=2}^{l_{\rm max}} \fr{(2l+1)\, \rho(l)^2}{l^2(l+1)^2\,\, C_{l}^{T T}} e^{-l^2/84^2}
\eeq
where  for $\mu_{\rm min}$  we take the PIXIE fiducial value $10^{-8}$  \cite{Ganc:2012ae}, and the sum in \eqref{StoN2} is carried up to $l_{\rm max} = 200$. To estimate the $S/N$, we would require the full knowledge of $C^{TT}_l$ in \eqref{CLTT} using the full transfer function $\Delta_l$. Taking these full transfer effects into account, \cite{Ganc:2012ae,Chluba:2016aln} found that the signal to noise ratio can be estimated as $40 \% $ of the result that one would obtain by adopting the SW limit $\rho_l \to 1$, $C^{TT,{\rm SW}}_l = 2\pi \mathcal{P}^{(0)}_{\mathcal{R}}/ (25\,l(l+1))$ \footnote{Note that the smallest scales we are interested corresponds to $l_{\rm max} =200$, so we can safely assume the scale indepedent part of the power spectrum in \eqref{CLTT} although it has non-trivial scale dependence (enhancement) at much smaller scales for inflationary scenarios we are interested in this work.} in \eqref{StoN2}. Keeping this in mind, we obtain 
\beq
\left(\fr{S}{N}\right) \simeq 0.35 \times 10^{-3} | f^{(\rm p)}_{\rm NL}  b^{(\rm SW)}_{(\rm pbh)} | \left(\fr{10^{-8}}{\mu_{\rm min}}\right).
\eeq
This result implies that $ | f^{(\rm p)}_{\rm NL}  b^{(\rm SW)}_{(\rm pbh)} | \gtrsim 2892$ should be observable for a PIXIE like experiment. On the other hand, for a spectrometer comparable to PRISM \cite{Andre:2013nfa} with $\mu_{\rm min} = 10^{-9}$,  $| f^{(\rm p)}_{\rm NL}  b^{(\rm SW)}_{(\rm pbh)} | \gtrsim 290$ is required for the detectability. At this point, it is worth mentioning that for a local type $f^{(\rm p)}_{\rm NL} \sim 10^{-2}$ at CMB scales as predicted by standard slow-roll backgrounds, such a signal would be challenging to detect. However, PBH forming inflationary models can in principle generate a more pronounced dip feature in the power spectrum (and hence in the bispectrum) (See \eg \cite{Ozsoy:2018flq,Ballesteros:2020qam}) compared to dip features we can capture using analytic formulas we study in this work. Based on our explorations in this paper, we anticipate that for such dramatic dip features, the overall magnitude of $b^{(\rm SW)}_{(\rm pbh)}$ can be in principle enhanced to compensate for the $f^{(\rm p)}_{\rm NL} \sim 0.01$ at CMB scales. A detailed analysis on the expected $b_{(\rm pbh)}$ in the aforementioned models require numerical work and is outside the scope of this work. 

On the other hand, considering the $2\sigma$ limits on the local type bispectrum by Planck \cite{Akrami:2019izv} at CMB scales: $-11.1 < f^{(\rm p)}_{\rm NL} < 9.3$ ($f_{\rm NL} = -0.9 \pm 5.1$,\,$\,\,68 \% {\rm CL}$) and the typical values of $|b^{(\rm SW)}_{(\rm pbh)}| \simeq {\rm a\,\, few}\,(10^2 - 10^3)$ that can be obtained in inflationary scenarios that can generate PBH populations (See figure \ref{fig:b}), we conclude that the influence of the scale dependent non-Gaussianity (see Section \ref{S3p3}) on $C^{\mu T}_l$, as due as the PBH formation mechanism, could be observable by PIXIE or PRISM for an interesting range of $k_{\rm dip}$ values.

\section{Discussion}
\label{sec-dis}

{In this work, we presented a method for probing inflationary scenarios
of PBH formation, using only CMB physics at relatively large
scales. We based our arguments on the characteristic properties of
the spectrum of curvature perturbation in single-field inflationary models that can generate a large population of PBHs. In these models, the curvature perturbation spectrum is characterized by a pronounced dip followed by a rapid growth towards a peak responsible for PBH formation. By making use of the gradient expansion formalism of \cite{Leach:2001zf},  we analytically computed the properties of the power spectrum and (for the first time) of the bispectrum around the dip position, which occurs at scales well larger than the peak. The bispectrum turns out to have a rich dependence on momenta, with a broad support spanning  different bispectrum  shapes. We found that when focussing
on isosceles and squeezed configurations, the bispectrum can be enhanced at the position of the dip; it also acquires a characteristic momentum dependence
that is controlled by  the underlying  inflationary mechanism. We proposed to probe such enhanced squeezed bispectrum  through the correlations
it induces between CMB $\mu$-distortions and CMB temperature fluctuations. We extended the methods first explored in \cite{Pajer:2012vz} to include the case of scale-dependent non-Gaussianity from PBH formation mechanisms, finding analytical expression for quantities controlling $\mu T$ correlations, and discussing how future CMB $\mu$-distortion experiments can test this observable. Interestingly, the method we propose would allow one to experimentally probe inflationary PBH scenarios using well-understood CMB physics at scales much larger than the peak of the spectrum, without considering non-linearities associated with PBH formation and evolution. In particular, owing the relation between relevant scales associated with $\mu$ distortions and the features present in the PBH forming inflationary scenarios, our findings are relevant for PBH masses within the $M_{(\rm pbh)} \sim 1-100 M_\odot$. This implies that $\mu T$ correlations can be considered as a useful tool to distinguish between astrophysical vs primordial origin of BHs.

Our work can be extended in several directions. From the phenomenological side, it would be interesting to explore more broadly the effects of non-Gaussianities in the region occurring around the dip, and {in particular to investigate} whether an enhanced trispectrum would lead to distinctive signals in the self-correlations of $\mu$-distortions   \cite{Pajer:2012vz}. Moreover, the ideas we pursued in this work can be complemented with more direct methods for constraining the slope of the scalar power spectrum with $\mu$-distortions \cite{Byrnes:2018txb,Unal:2020mts}, that rely on the knowledge of the growth rate of the spectrum towards the peak and hence control different ranges of scales.  {On } the theoretical side, it would be interesting to have a better understanding of possible consistency relations connecting the features of the spectrum such as dips and peaks \cite{Tasinato:2020vdk}, and their consequences for the bispectrum. Finally, {extending the ideas we presented in this work to the multiple scalar field case appear as another interesting venue to be explored, in particular to make a comparison with single-field results}.  Then a pronounced dip feature in the power spectrum might be missing \cite{Garcia:2020mwi,Palma:2020ejf,Fumagalli:2020adf,Fumagalli:2020nvq} (but see  \cite{Braglia:2020taf}). Similarly, in models that utilize axion gauge-field dynamics during inflation to enhance the curvature perturbation, such a future is generically not present\cite{Garcia-Bellido:2016dkw,Ozsoy:2020ccy} (See however \cite{Ozsoy:2020kat}). On the other hand, still much work is needed to clarify the properties of the spectrum in  this context. 
}

\subsection*{Acknowledgments}
It is a pleasure to thank Enrico Pajer for comments on the manuscript. 
The work of O\"O is supported by the European Structural and Investment Funds and the Czech Ministry of Education, Youth and Sports (Project CoGraDS-CZ.02.1.01/0.0/0.0/15003/0000437). GT is partially funded by the STFC grant ST/T000813/1. 
\begin{appendix}
\section{The curvature perturbation $\mathcal{R}_k$ and fractional velocity $v_{\mathcal{R}}$}\label{AppA}

The enhancement factor $\alpha_k$ in \eqref{ar}, \eqref{ai} and the eqs.  \eqref{psf1} and \eqref{3PT} implies that we need an expression for $\mathcal{R}(\tau_k)$ and the fractional velocity $v_{\mathcal{R}}$ in \eqref{fr} for being able to determine the spectral behavior of power and bispectrum of $\mathcal{R}_k$.  In this appendix, we therefore aim to derive an expression for $\mathcal{R}(\tau_k)$ and $v_{\mathcal{R}}$ for the two phase model we study in the main text. For this purpose, we resort to Mukhanov-Sasaki equation for the canonically normalized variable $Q_k(\tau) \equiv z(\tau) \mathcal{R}_k(\tau)$,
\beq\label{MS}
Q_k''+\left(k^2 - \fr{z''}{z}\right)Q_k = 0,
\eeq
where 
\beq
\fr{z''}{z} =(a H)^2\left[2-\epsilon+\fr{3}{2}\eta+\fr{1}{4}\eta^2-\fr{1}{2}\epsilon\eta+\fr{1}{2}\fr{\dot{\eta}}{H}\right], 
\eeq
which is exact to all orders in slow-roll parameters. For constant values of slow-roll parameters $\epsilon,\eta$, and assuming $\epsilon \ll\ll1$ (such that $H$ is approximately constant as we assume), the exact solution for $Q_k$ in terms of the Hankel functions is given by 
\beq\label{MSgs}
Q_k = A \sqrt{-\tau} H^{(1)}_{\nu}(-k\tau) + B \sqrt{-\tau} H^{(2)}_{\nu}(-k\tau).
\eeq
where
\beq
\nu^2 \simeq \fr{9}{4}+\fr{3}{2}\eta+\fr{1}{4}\eta^2 = \left(\fr{3+\eta}{2}\right)^2.
\eeq
For the initial slow-roll phase ($\eta_{\rm sr}=0$ $\to$ $\nu = 3/2$), requiring that all modes are in their Bunch-Davies vacuum for $-k\tau \to \infty$ in \eqref{MSgs}, the curvature perturbation  is then given by 
  
\beq\label{cpsr}
\mathcal{R}^{\rm sr}_k = \fr{i H}{\Mp} \fr{e^{-ik\tau}}{\sqrt{4\epsilon_{\rm sr} k^3}}~(1+ik\tau),\quad\quad\quad \tau_k/\tau_0 > 1.
\eeq
where we have used $z = (-H\tau)^{-1} \sqrt{2\epsilon_{\rm sr}} \Mp$. The solution above immediately implies
\beq\label{fvsr}
\fr{\mathcal{R}_k'}{3\mathcal{H}\mathcal{R}_k} =-\fr{(-k\tau)^2+i(-k\tau)^3}{3(1+(-k\tau)^2)}.
\eeq
This result makes it clear why the curvature perturbation settles to a constant solution shortly after  horizon exit in standard slow-roll inflation, which can be understood in the $-k\tau \to 0$ limit of  eq. \eqref{fvsr}. For our purposes, we are interested in the expression in \eqref{fvsr} evaluated at the initial time $\tau=\tau_k$ at around horizon crossing. With this in mind, we split the fractional velocity into a real and imaginary part as,
\beq\label{vsr}
v^{R}_{\mathcal{R}}(c_k) = -\fr{c_k^2}{3(1+c_k^2)},\quad
v^{I}_{\mathcal{R}}(c_k)  = -\fr{c_k^3}{3(1+c_k^2)},\quad\quad \tau_k/\tau_0 > 1,
\eeq
where we defined a positive number $-k\tau_k = k/\mathcal{H}_k \equiv c_k \leq 1$ to identify the size of the each mode with respect to the horizon size at the initial time, \ie at $\tau = \tau_k$. It is clear from this expression that the imaginary part of $v_{\mathcal{R}}$ includes an extra factor of $c_k$ compared to the real part. We note that unless $c_k =1$, this translates into an extra suppression for the imaginary part of the fractional velocity.

On the other hand, using \eqref{cpsr}, the power spectrum evaluated at around horizon crossing is given by
\beq\label{pstauk}
\mathcal{P}_{\mathcal{R}}(\tau_k) = \fr{k^3}{2\pi^2}|\mathcal{R}_k(\tau_k)|^2 = \fr{H^2}{8 \pi^2\epsilon_{\rm sr}\Mp^2} \left(1 + c_k^2\right),\quad\quad\quad \tau_k/\tau_0 > 1.
\eeq 

Next, we need to develop an expression for $\mathcal{R}(\tau_k)$ and the fractional velocity continuous through the transition at $\tau =\tau_0$. For this purpose, we  use a matching procedure for $\mathcal{R}_k$ and its derivative between the initial slow-roll era, \ie \eqref{cpsr} to a general solution during the non-attractor stage:
\beq\label{cpna}
\mathcal{R}_k = \fr{i H}{\Mp} \fr{(\tau/\tau_0)^{\eta_{\rm c}/2}}{\sqrt{4\epsilon_{\rm sr} k^3}}(-k\tau)^{3/2}\left[c_1 H^{(1)}_{\nu}(-k\tau)+c_2 H^{(2)}_{\nu}(-k\tau)\right], \quad\quad \tau_k/\tau_0 < 1.
\eeq
where $\nu = (3+\eta_{\rm c})/2$. Matching $\mathcal{R}_k$ and $\mathcal{R}'_k$ at the transition $\tau=\tau_0$ using \eqref{cpsr} and \eqref{cpna}, we obtain
\bea
c_1 &=& y_0^{3/2}e^{iy_0}\fr{y_0\left(H^{(2)}_\nu(y_0)+iH^{(2)}_{\nu-1}(y_0)\right)-H^{(2)}_{\nu-1}(y_0)}{H^{(1)}_{\nu-1}(y_0)H^{(2)}_{\nu}(y_0)-H^{(1)}_{\nu}(y_0)H^{(2)}_{\nu-1}(y_0)},\\\nn\\
c_2 &= & y_0^{3/2}e^{iy_0}\fr{(y_0)\left(H^{(1)}_\nu(y_0)+iH^{(1)}_{\nu-1}(y_0)\right)-H^{(1)}_{\nu-1}(y_0)}{H^{(1)}_{\nu}(y_0)H^{(2)}_{\nu-1}(y_0)-H^{(2)}_{\nu}(y_0)H^{(1)}_{\nu-1}(y_0)}.
\eea
where we defined $y\equiv-k\tau$. Then for $\tau_k / \tau_0 < 1$, the real and the imaginary part of the fractional velocity is given by 
\bea
\label{fvintr}v^{R}_{\mathcal{R}}(\tau) &=&-\fr{y}{3}\left[\fr{f_1f_3-y_0\left(f_1f_4+f_2f_3\right)+y_0^2\left(f_1f_3+f_2f_4\right)}{f_3^2-2y_0f_3f_4+y_0^2\left(f_3^2+f_4^2\right)}\right],\\\nn\\
\label{fvinti}v^{I}_{\mathcal{R}}(\tau) &=&-\fr{y}{3}\left[\fr{y_0^2\left(f_1f_4-f_2f_3\right)}{f_3^2-2y_0f_3f_4+y_0^2\left(f_3^2+f_4^2\right)}\right],
\eea
where we define the functions $f_\alpha = f_\alpha(y,y_0,\nu)$, $\alpha=1,2,3,4$ in terms of the Bessel function of the first and second kind as
\bea
f_1(y,y_0,\nu) &=& J_{\nu-1}(y_)Y_{\nu-1}(y)-Y_{\nu-1}(y_0)J_{\nu-1}(y),\\
f_2(y,y_0,\nu) &=& J_{\nu}(y_0)Y_{\nu-1}(y)-Y_{\nu}(y_0)J_{\nu-1}(y),
\eea
noting $f_4 = f_1(y,y_0,\nu+1), f_3 = -f_2(y_0,y,\nu)$. Using these expressions, power spectrum evaluated at $\tau = \tau_k$ for modes that leave the horizon during the non-attractor phase is given by
\beq\label{pstaukna}
\mathcal{P}_{\mathcal{R}}(\tau_k)  = \fr{H^2}{8\pi^2\epsilon_{\rm sr}\Mp^2}\left(\fr{\tau_k}{\tau_0}\right)^{2\nu}\left[\fr{f_3^2-2y_0f_3f_4+y_0^2(f_3^2+f_4^2)}{f_3(y_0,y_0,\nu)^2}\right]_{\tau= \tau_k},\quad\quad\quad \tau_k/\tau_0 < 1.
\eeq 

The continuity of the real and the imaginary part of the fractional velocity can be confirmed explicitly from the eqs in \eqref{fvintr} and \eqref{fvinti} as they reduce to their slow-roll counterparts provided in \eqref{vsr} at the transition point $\tau_k=\tau_0$.

\section{The functions $D(\tau_k)$,  $F_k(\tau_k)$}\label{AppB} 

In this appendix, we present the details on the calculation of the integrals associated with the functions $D(\tau_k)$,$F(\tau_k)$, 
in the background model given in eq. \eqref{zsol1}. For this purpose, we define $x\equiv \tau/\tau_0$ to re-parametrize the background pump field as
\beq\label{zsol1a}
z(\tau)= 
\left\{    
 \begin{array}{rl}
&
z_0 ~x^{-1}, \hskip1.6cm x \geq 1,
\\
&
z_0\, x^{-(\eta_{\rm c}+2)/2}\hskip0.8cm x_f \leq x \leq 1.
   \end{array}\right. \,
\eeq 

For $x_k >1$, \ie for modes leaving the horizon during the slow-roll era, the functions $D(\tau_k)$ and $F(\tau_k)$ were calculated in \cite{Ozsoy:2019lyy} and found to be
\begin{align}
\label{dafa1}
 D(\tau_k) &= 1 - \fr{3}{(\eta_{\rm c}+3)} \left[e^{-(\eta_{\rm c}+3)\Delta N}+\fr{\eta_{\rm c}}{3}\right]x_k^{-3},\\\nn
\fr{F(\tau_k)}{\tau_0^2} &= -\left[\fr{ \eta_{\rm c}\,e^{-(\eta_{\rm c}+3)\dn}}{(\eta_{\rm c}+3)(\eta_{\rm c}+1)}+\fr{\eta_{\rm c}}{2(\eta_{\rm c}+3)}\right]+\left[\fr{e^{-(\eta_{\rm c}+3)\dn}}{(\eta_{\rm c}+3)}+\fr{\eta_{\rm c}}{3(\eta_{\rm c}+3)}\right]x_k^{-1}+\fr{x_k^2}{6}-\fr{e^{-2\dn}}{2(\eta_{\rm c}+1)},
\end{align}
where $\tau_f/\tau_0 \equiv x_f = e^{-\dn}$ with $\Delta N$ denoting the duration of the non-attractor era.
Next we focus on $D(\tau_k)$ and $F(\tau_k)$ for modes that leave the horizon during the non-attractor era, \ie $x_k < 1$. In this case, integrals defined in \eqref{Dint} and \eqref{Fint} are much simpler to evaluate which require only the behavior of the pump field during the non-attractor era, \ie the second line in \eqref{zsol1a}. Proceeding in this way, for modes that exit during the non-attractor era, $x_k < 1$, we obtain the following expressions 
\begin{align}\label{dafa2}
\nn D(\tau_k) &= -\fr{3}{\eta_{\rm c}+3}\left[e^{-(\eta_{\rm c}+3)\dn}\,x_k^{-(\eta_{\rm c}+3)}+1\right],\\
\fr{F(\tau_k)}{\tau_0^2} &= \fr{e^{-(\eta_{\rm c}+3)\dn}}{(\eta_{\rm c}+1)(\eta_{\rm c}+3)}\,x_k^{-(\eta_{\rm c}+1)} + \fr{x_k^2}{2(\eta_{\rm c}+3)}-\fr{e^{-2\Delta N}}{2(\eta_{\rm c}+1)}.
\end{align}
Denoting $x_k = c_k (k/\mathcal{H}_0)^{-1}$, the wave-number dependence of these functions can be parametrized as
\begin{align}\label{dafa1f}
D(\tau_k) &\equiv  \cc_{0}^{D} + \cc_{3}^D ~\left(\fr{k}{\mathcal{H}_0}\right)^{3},\\\nn
\fr{F(\tau_k)}{\tau_0^2} &\equiv \cc_{-2}^F ~\left(\fr{k}{\mathcal{H}_0}\right)^{-2} + \cc_{0}^{F}+ \cc_{1}^F ~\left(\fr{k}{\mathcal{H}_0}\right),\quad\quad\quad\quad\quad\fr{k}{\mathcal{H}_0}< c_k
\end{align}
and 
\begin{align}\label{dafa2f}
D(\tau_k) &\equiv   \tilde{\cc}_{\eta_{\rm c}+3}^D ~\left(\fr{k}{\mathcal{H}_0}\right)^{\eta_{\rm c}+3} + \tilde{\cc}_{0}^{D} \\\nn
\fr{F(\tau_k)}{\tau_0^2} &\equiv \tilde{\cc}_{\eta_{\rm c}+1}^F ~\left(\fr{k}{\mathcal{H}_0}\right)^{\eta_{\rm c}+1}+\tilde{\cc}_{-2}^F ~\left(\fr{k}{\mathcal{H}_0}\right)^{-2} + \tilde{\cc}_{0}^{F},\quad\quad\quad\quad c_k < \fr{k}{\mathcal{H}_0} < e^{\Delta N}
\end{align}
where the coefficients $\mathcal{C}, \tilde{\mathcal{C}}$ are functions of the parameter set $\{ c_k,\eta_{\rm c}, \Delta N \}$ as can be understood from \eqref{dafa1}. Note that for modes that exit the horizon during the non-attractor era, the formulas in \eqref{dafa2f} are valid up to a maximum wave-number $k/\mathcal{H}_0$ corresponding to the mode that exits the horizon at the end of the non-attractor era which obeys $k \tau_f = 1$.

It is worth emphasizing that the coefficients $\cc$ and $\tilde{\cc}$ that multiply the $k$ dependent terms can be organized in a hierarchal way in powers (determined by $\etc$) of $a(\tau_f)/a(\tau_0) = e^{\Delta N}$ where $\Delta N$ is the duration of non-attractor era in number of e-folds. This result reflects the fact that modes that leave during the slow-roll and in the early stages of non-attractor era are enhanced due to the slow-roll violation $\eta_{\rm c}\leq -6$. 
\section{Consistency condition}\label{AppC}
In this appendix, our aim is to show that the formulas we derived for the non-linearity parameter in \eqref{fnlf} reduce to the standard expression implied by the consistency condition for vanilla slow-roll models. For this purpose, we will focus on the mode equation \eqref{CPE} of the comoving curvature perturbation in Fourier space. 
It is a well known fact that in a standard slow-roll background, the growing mode solution of eq. \eqref{CPE} is conserved on super horizon scales. This can be readily seen from the formal integral solution of \eqref{CPE}, which can be written up to order $\mathcal{O}(k^2)$ for small but finite wave-numbers as 
\beq\label{sme}
\mathcal{R}_k(\tau)\simeq \mathcal{R}^{(0)} \left[ 1 + \mathcal{C}_2 \int_{\tau_k}^{\tau} \fr{\mathrm{d} \tau'}{z^{2}(\tau')}-k^{2} \int_{\tau_k}^{\tau} \frac{\mathrm{d} \tau'}{z^{2}(\tau')} \int_{\tau_k}^{\tau'} \d \tau'' z^{2}(\tau'') \right] ,
\eeq
where we obtain the last term by solving iteratively the inhomogeneous part of eq. \eqref{CPE} using the leading growing mode which we identify as $\mathcal{R}_k(\tau_k)=\mathcal{R}^{(0)}$. The constant behavior of $\mathcal{R}_{k}$ shortly after its scale crosses the horizon can be readily seen from the solution \eqref{sme}, by realizing  that
 -- in a slow-roll background where  $z \propto (-\tau)^{-1}$ --
 the second and the third term in \eqref{sme}  decay respectively as
   $(-\tau)^3$ and $(-\tau)^2$   in the late time limit $-\tau \to 0$. Therefore, in a slow-roll background we can immediately identify the second and third term in \eqref{sme} as the decaying modes. In fact, the standard decaying mode is given by the last term as it decays slowly,  \ie $\propto (-k\tau)^2$, compared to the second. Notice that the second and the third term are proportional to the functions we defined as $D(\tau)$ and $F(\tau)$ within the gradient expansion formalism we are undertaken. In the language of the main text, the discussion above suggests that in an always slow-roll background, we can neglect the terms proportional to $D(\tau)$ compared to the term proportional to $F(\tau)$ for a sufficiently late time $\tau_k$ shortly after horizon crossing. Therefore, the power spectrum at late times can be well approximated by 
\beq\label{pssr}
\mathcal{P}_\mathcal{R}(\tau_*, k) = |\alpha_k|^2 \mathcal{P}_{\mathcal{R}}(\tau_k) \simeq  |1 - F(\tau_k)k^2|^2 \mathcal{P}^{(0)}_{\mathcal{R}},
\eeq
where $\mathcal{P}^{(0)}_{\mathcal{R}} = k^3 |\mathcal{R}^{(0)}|^2/(2\pi^2) = H^2/(8\pi^2 \epsilon_{\rm sr} \Mp^2) $ is the scale invariant power spectrum evaluated at $\tau_k$ assuming a constant $H$ with $\epsilon_{\rm sr} \ll 1$. From \eqref{pssr}, we can relate spectral index to the $k$ dependent part of the $\alpha_k$ as
\beq\label{sindex}
\fr{\d \ln \mathcal{P}_\mathcal{R}(\tau_*, k)}{\d \ln k} \equiv n_s - 1 = \fr{2}{|\alpha_k|}\fr{\d |\alpha_k|}{\d \ln k} \approx -4 F(\tau_k) k^2,
\eeq
where $F(\tau_k) = \tau_k^2/6$ in a slow-roll background with $z \propto (-\tau)^{-1}$ and used the fact that there is no enhancement in this scenario, \ie $|\alpha_k| \to 1$. Taking the squeezed limit $k_1\simeq k_2 = q\gg k_3$ of the expression in \eqref{fnlf} then gives 
\beq
 f_{\rm NL}\left(q, q, k_{3}\right) = \fr{5}{12}\fr{4 F(\tau_{q})q^2 }{\alpha_{q} \alpha^{*}_{k_{3}}} \simeq \fr{5}{12} (1-n_s)
\eeq
where we used \eqref{sindex} by noting $|\alpha_{k_3}| \simeq 1$ as $k_3 \to 0$ and $|\alpha_q| \simeq 1$ for a slow-roll background as before. It is worth to point out that by virtue of the gradient expansion formalism we undertake, $1-n_s \approx 2 c_k^2 / 3 \ll 1$ at leading order in the small parameter $c_k^2$ with $c_k = -k\tau_k < 1$ which is consistent with the identification that all the modes we consider are already outside the horizon at the initial time $\tau_k$.


\section{$f^{\rm eff}_{\rm NL}$ in the squeezed limit}\label{AppD}
Here, we would like to derive the squeezed limit of the effective non-linearity parameter defined in \eqref{fnleffna} for modes that leave the horizon before the background transitions to the non-attractor regime. Using the formulas we derived for the functions $D(\tau_k),F(\tau_k)$ and the definitions of the enhancement factor $\alpha_k$ (See \eg eqs. \eqref{ar} and \eqref{ai}), we can derive and expression for $f^{\rm eff}_{\rm NL}$ \eqref{fnleffna} in terms of an expansion over the small ratio $k / q \ll 1$ in the squeezed limit. In this way, up to next to leading order in $k/q$, we obtain
\beq\label{fnleffapp}
f^{\rm eff}_{\rm NL} \simeq \fr{5}{12} \left\{4 F(\tau_q) q^2\bigg[ \alpha^{R}_q\left(1 + v^{R}_{\mathcal{R}}(c_{k}) - \mathcal{C}^{F}_{-2}(c_{k})\right)+\alpha^{I}_q\,v^{I}_{\mathcal{R}}(c_{k})\bigg] +5 |\alpha_{q}|^2 \mathcal{C}^{F}_{-2}(c_{k})\,\fr{k}{q}+\mathcal{O}\left(\fr{k^2}{q^2}\right)\right\},
\eeq
where $v_\mathcal{R}$'s are defined as in \eqref{vsr} and the background dependent coefficients $\mathcal{C}^{F}_{-2} = \mathcal{C}^{F}_{-2}(c_k)$ can be extracted from \eqref{dafa1} and \eqref{dafa1f}. Similarly, in order to obtain an explicit expression in terms of the hard momenta $q$, we further dissect the expression in \eqref{fnleffapp} using the formulas for $\alpha_q$ and $F(\tau_q)$ to re-write the effective non-linearity parameter as 
\beq\label{fnleffappf}
\bar{f}^{\rm eff}_{\rm NL} \equiv \fr{f^{\rm eff}_{\rm NL}}{f^{(0)}_{\rm NL}} \simeq 1 +\sum^{6}_{n=2} \fr{c^{(n)}_{{\rm NL}}(\eta_{\rm c},\Delta N,c_q)}{c^{(0)}_{{\rm NL}}(c_q)}\left(\fr{q}{\mathcal{H}_0}\right)^{n},
\eeq
where $f^{(0)}_{\rm NL} = 5 c^{(0)}_{{\rm NL}}(c_q) /3$ is the ‘‘initial" value of the $f^{\rm eff}_{\rm NL}$ we define in the large scale tail of the squeezed limit, \ie $q\gg k$ as $q \to 0$ and we used the fact that we universally utilize a small number for all the modes we study assuming $c_{k} = c_q = \textrm{constant}< 1$. Recall that the condition $c_k = -k \tau_k < 1$ serve for the purpose to ensure that each mode labelled by a wave-number $k$ are outside the horizon for an appropriately chosen time $\tau_k$. The coefficients $c^{(n)}_{\rm NL}$ that appear in \eqref{fnleffappf} can be written in terms of the coefficients $\mathcal{C}$ of the functions $F,D$ in Appendix \ref{AppB} and the fractional velocities $v^{R}_{\mathcal{R}}, v^{I}_{\mathcal{R}}$ we defined in Appendix \ref{AppA} as
\begin{align}
\label{d3}c^{(0)}_{{\rm NL}} &\equiv \left[(1+v^{R}_{\mathcal{R}}-\mathcal{C}^{F}_{-2})^2 + (v^{I}_{\mathcal{R}})^2\right] \mathcal{C}^{F}_{-2},\\
c^{(2)}_{{\rm NL}}  &\equiv  \mathcal{C}^{F}_{0}\left[\fr{c^{(0)}_{{\rm NL}}}{\mathcal{C}^{F}_{-2}} - (1+v^{R}_{\mathcal{R}}-\mathcal{C}^{F}_{-2}) \mathcal{C}^{F}_{-2}\right],\\
c^{(3)}_{{\rm NL}}  &\equiv \fr{c^{(0)}_{{\rm NL}}}{\mathcal{C}^{F}_{-2}}\, \mathcal{C}^{F}_{1} + (1+v^{R}_{\mathcal{R}}-\mathcal{C}^{F}_{-2})\left[v^{R}_{\mathcal{R}}\,\mathcal{C}^{D}_{3}-\mathcal{C}^{F}_{1}\right] \mathcal{C}^{F}_{-2} + (v^{I}_{\mathcal{R}})^2 \mathcal{C}^{D}_{3}\, \mathcal{C}^{F}_{-2},\\
c^{(4)}_{{\rm NL}}  &\equiv - (1+v^{R}_{\mathcal{R}}-\mathcal{C}^{F}_{-2})\, ( \mathcal{C}^{F}_{0} )^2,\\
c^{(5)}_{{\rm NL}}  &\equiv  (1+v^{R}_{\mathcal{R}}-\mathcal{C}^{F}_{-2})\left[v^{R}_{\mathcal{R}}\,\mathcal{C}^{D}_{3}-2\,\mathcal{C}^{F}_{1}\right] \mathcal{C}^{F}_{0} + (v^{I}_{\mathcal{R}})^2 \mathcal{C}^{D}_{3}\, \mathcal{C}^{F}_{0},\\
c^{(6)}_{{\rm NL}}  &\equiv (1+v^{R}_{\mathcal{R}}-\mathcal{C}^{F}_{-2})\left[v^{R}_{\mathcal{R}}\,\mathcal{C}^{D}_{3}-\mathcal{C}^{F}_{1}\right] \mathcal{C}^{F}_{1} + (v^{I}_{\mathcal{R}})^2 \mathcal{C}^{D}_{3}\, \mathcal{C}^{F}_{1},\label{d8}
\end{align}
where $\{\eta_{\rm c},\Delta N, c_q\}$ dependence of these expressions should be understood. 
\end{appendix}

\addcontentsline{toc}{section}{References}
\bibliographystyle{utphys}

\bibliography{paper2}

\providecommand{\href}[2]{#2}\begingroup\raggedright\begin{thebibliography}{10}

\bibitem{Hawking:1971ei}
S.~Hawking, ``{Gravitationally collapsed objects of very low mass},''
{\em Mon. Not. Roy. Astron. Soc.} {\bfseries 152} (1971) 75.

\bibitem{Carr:1974nx}
B.~J. Carr and S.~W. Hawking, ``{Black holes in the early Universe},''
{\em Mon. Not. Roy. Astron. Soc.} {\bfseries 168} (1974) 399--415.

\bibitem{Carr:1975qj}
B.~J. Carr, ``{The Primordial black hole mass spectrum},''
\href{http://dx.doi.org/10.1086/153853}{{\em Astrophys. J.} {\bfseries 201}
  (1975) 1--19}.

\bibitem{Khlopov:2008qy}
M.~Y. Khlopov, ``{Primordial Black Holes},''
  \href{http://dx.doi.org/10.1088/1674-4527/10/6/001}{{\em Res. Astron.
  Astrophys.} {\bfseries 10} (2010) 495--528},
  \href{http://arxiv.org/abs/0801.0116}{{\ttfamily arXiv:0801.0116
  [astro-ph]}}.

\bibitem{Belotsky:2014kca}
K.~M. Belotsky, A.~D. Dmitriev, E.~A. Esipova, V.~A. Gani, A.~V. Grobov, M.~Y.
  Khlopov, A.~A. Kirillov, S.~G. Rubin, and I.~V. Svadkovsky, ``{Signatures of
  primordial black hole dark matter},''
  \href{http://dx.doi.org/10.1142/S0217732314400057}{{\em Mod. Phys. Lett. A}
  {\bfseries 29} no.~37, (2014) 1440005},
  \href{http://arxiv.org/abs/1410.0203}{{\ttfamily arXiv:1410.0203
  [astro-ph.CO]}}.

\bibitem{Ivanov:1994pa}
P.~Ivanov, P.~Naselsky, and I.~Novikov, ``{Inflation and primordial black holes
  as dark matter},''
\href{http://dx.doi.org/10.1103/PhysRevD.50.7173}{{\em Phys. Rev.} {\bfseries
  D50} (1994) 7173--7178}.

\bibitem{GarciaBellido:1996qt}
J.~Garcia-Bellido, A.~D. Linde, and D.~Wands, ``{Density perturbations and
  black hole formation in hybrid inflation},''
  \href{http://dx.doi.org/10.1103/PhysRevD.54.6040}{{\em Phys. Rev.} {\bfseries
  D54} (1996) 6040--6058},
\href{http://arxiv.org/abs/astro-ph/9605094}{{\ttfamily arXiv:astro-ph/9605094
  [astro-ph]}}.

\bibitem{Germani:2018jgr}
C.~Germani and I.~Musco, ``{Abundance of Primordial Black Holes Depends on the
  Shape of the Inflationary Power Spectrum},''
  \href{http://dx.doi.org/10.1103/PhysRevLett.122.141302}{{\em Phys. Rev.
  Lett.} {\bfseries 122} no.~14, (2019) 141302},
\href{http://arxiv.org/abs/1805.04087}{{\ttfamily arXiv:1805.04087
  [astro-ph.CO]}}.

\bibitem{Carr:2016drx}
B.~Carr, F.~Kuhnel, and M.~Sandstad, ``{Primordial Black Holes as Dark
  Matter},'' \href{http://dx.doi.org/10.1103/PhysRevD.94.083504}{{\em Phys.
  Rev.} {\bfseries D94} no.~8, (2016) 083504},
\href{http://arxiv.org/abs/1607.06077}{{\ttfamily arXiv:1607.06077
  [astro-ph.CO]}}.

\bibitem{Sasaki:2018dmp}
M.~Sasaki, T.~Suyama, T.~Tanaka, and S.~Yokoyama, ``{Primordial black
  holes—perspectives in gravitational wave astronomy},''
  \href{http://dx.doi.org/10.1088/1361-6382/aaa7b4}{{\em Class. Quant. Grav.}
  {\bfseries 35} no.~6, (2018) 063001},
\href{http://arxiv.org/abs/1801.05235}{{\ttfamily arXiv:1801.05235
  [astro-ph.CO]}}.

\bibitem{Carr:2020xqk}
B.~Carr and F.~Kuhnel, ``{Primordial Black Holes as Dark Matter: Recent
  Developments},''
  \href{http://dx.doi.org/10.1146/annurev-nucl-050520-125911}{{\em Ann. Rev.
  Nucl. Part. Sci.} {\bfseries 70} (2020) 355--394},
  \href{http://arxiv.org/abs/2006.02838}{{\ttfamily arXiv:2006.02838
  [astro-ph.CO]}}.

\bibitem{Green:2020jor}
A.~M. Green and B.~J. Kavanagh, ``{Primordial Black Holes as a dark matter
  candidate},'' \href{http://dx.doi.org/10.1088/1361-6471/abc534}{{\em J. Phys.
  G} {\bfseries 48} no.~4, (2021) 4},
  \href{http://arxiv.org/abs/2007.10722}{{\ttfamily arXiv:2007.10722
  [astro-ph.CO]}}.

\bibitem{Byrnes:2018txb}
C.~T. Byrnes, P.~S. Cole, and S.~P. Patil, ``{Steepest growth of the power
  spectrum and primordial black holes},''
  \href{http://dx.doi.org/10.1088/1475-7516/2019/06/028}{{\em JCAP} {\bfseries
  1906} no.~06, (2019) 028},
\href{http://arxiv.org/abs/1811.11158}{{\ttfamily arXiv:1811.11158
  [astro-ph.CO]}}.

\bibitem{Tasinato:2020vdk}
G.~Tasinato, ``{An analytic approach to non-slow-roll inflation},''
  \href{http://dx.doi.org/10.1103/PhysRevD.103.023535}{{\em Phys. Rev. D}
  {\bfseries 103} no.~2, (2021) 023535},
  \href{http://arxiv.org/abs/2012.02518}{{\ttfamily arXiv:2012.02518
  [hep-th]}}.

\bibitem{Leach:2001zf}
S.~M. Leach, M.~Sasaki, D.~Wands, and A.~R. Liddle, ``{Enhancement of
  superhorizon scale inflationary curvature perturbations},''
  \href{http://dx.doi.org/10.1103/PhysRevD.64.023512}{{\em Phys. Rev.}
  {\bfseries D64} (2001) 023512},
\href{http://arxiv.org/abs/astro-ph/0101406}{{\ttfamily arXiv:astro-ph/0101406
  [astro-ph]}}.

\bibitem{Ozsoy:2019lyy}
O.~\"Ozsoy and G.~Tasinato, ``{On the slope of the curvature power spectrum in
  non-attractor inflation},''
  \href{http://dx.doi.org/10.1088/1475-7516/2020/04/048}{{\em JCAP} {\bfseries
  04} (2020) 048}, \href{http://arxiv.org/abs/1912.01061}{{\ttfamily
  arXiv:1912.01061 [astro-ph.CO]}}.

\bibitem{Pajer:2012vz}
E.~Pajer and M.~Zaldarriaga, ``{A New Window on Primordial non-Gaussianity},''
  \href{http://dx.doi.org/10.1103/PhysRevLett.109.021302}{{\em Phys. Rev.
  Lett.} {\bfseries 109} (2012) 021302},
  \href{http://arxiv.org/abs/1201.5375}{{\ttfamily arXiv:1201.5375
  [astro-ph.CO]}}.

\bibitem{Mylova:2018yap}
M.~Mylova, O.~Özsoy, S.~Parameswaran, G.~Tasinato, and I.~Zavala, ``{A new
  mechanism to enhance primordial tensor fluctuations in single field
  inflation},'' \href{http://dx.doi.org/10.1088/1475-7516/2018/12/024}{{\em
  JCAP} {\bfseries 1812} no.~12, (2018) 024},
\href{http://arxiv.org/abs/1808.10475}{{\ttfamily arXiv:1808.10475 [gr-qc]}}.

\bibitem{Ozsoy:2019slf}
O.~Ozsoy, M.~Mylova, S.~Parameswaran, C.~Powell, G.~Tasinato, and I.~Zavala,
  ``{Squeezed tensor non-Gaussianity in non-attractor inflation},''
  \href{http://dx.doi.org/10.1088/1475-7516/2019/09/036}{{\em JCAP} {\bfseries
  1909} no.~09, (2019) 036},
\href{http://arxiv.org/abs/1902.04976}{{\ttfamily arXiv:1902.04976 [hep-th]}}.

\bibitem{Mukhanov:2005sc}
V.~Mukhanov, {\em {Physical Foundations of Cosmology}}.
\newblock Cambridge University Press, Oxford, 2005.
\newblock
\url{http://www-spires.fnal.gov/spires/find/books/www?cl=QB981.M89::2005}.
\newblock

\bibitem{Takamizu:2010xy}
Y.-i. Takamizu, S.~Mukohyama, M.~Sasaki, and Y.~Tanaka, ``{Non-Gaussianity of
  superhorizon curvature perturbations beyond $\delta$ N formalism},''
  \href{http://dx.doi.org/10.1088/1475-7516/2010/06/019}{{\em JCAP} {\bfseries
  1006} (2010) 019},
\href{http://arxiv.org/abs/1004.1870}{{\ttfamily arXiv:1004.1870
  [astro-ph.CO]}}.

\bibitem{Motohashi:2017kbs}
H.~Motohashi and W.~Hu, ``{Primordial Black Holes and Slow-Roll Violation},''
  \href{http://dx.doi.org/10.1103/PhysRevD.96.063503}{{\em Phys. Rev.}
  {\bfseries D96} no.~6, (2017) 063503},
\href{http://arxiv.org/abs/1706.06784}{{\ttfamily arXiv:1706.06784
  [astro-ph.CO]}}.

\bibitem{Garcia-Bellido:2017mdw}
J.~Garcia-Bellido and E.~Ruiz~Morales, ``{Primordial black holes from single
  field models of inflation},''
  \href{http://dx.doi.org/10.1016/j.dark.2017.09.007}{{\em Phys. Dark Univ.}
  {\bfseries 18} (2017) 47--54},
\href{http://arxiv.org/abs/1702.03901}{{\ttfamily arXiv:1702.03901
  [astro-ph.CO]}}.

\bibitem{Ezquiaga:2017fvi}
J.~M. Ezquiaga, J.~Garcia-Bellido, and E.~Ruiz~Morales, ``{Primordial Black
  Hole production in Critical Higgs Inflation},''
  \href{http://dx.doi.org/10.1016/j.physletb.2017.11.039}{{\em Phys. Lett.}
  {\bfseries B776} (2018) 345--349},
\href{http://arxiv.org/abs/1705.04861}{{\ttfamily arXiv:1705.04861
  [astro-ph.CO]}}.

\bibitem{Ballesteros:2017fsr}
G.~Ballesteros and M.~Taoso, ``{Primordial black hole dark matter from single
  field inflation},'' \href{http://dx.doi.org/10.1103/PhysRevD.97.023501}{{\em
  Phys. Rev.} {\bfseries D97} no.~2, (2018) 023501},
\href{http://arxiv.org/abs/1709.05565}{{\ttfamily arXiv:1709.05565 [hep-ph]}}.

\bibitem{Hertzberg:2017dkh}
M.~P. Hertzberg and M.~Yamada, ``{Primordial Black Holes from Polynomial
  Potentials in Single Field Inflation},''
  \href{http://dx.doi.org/10.1103/PhysRevD.97.083509}{{\em Phys. Rev.}
  {\bfseries D97} no.~8, (2018) 083509},
\href{http://arxiv.org/abs/1712.09750}{{\ttfamily arXiv:1712.09750
  [astro-ph.CO]}}.

\bibitem{Cicoli:2018asa}
M.~Cicoli, V.~A. Diaz, and F.~G. Pedro, ``{Primordial Black Holes from String
  Inflation},'' \href{http://dx.doi.org/10.1088/1475-7516/2018/06/034}{{\em
  JCAP} {\bfseries 1806} (2018) 034},
\href{http://arxiv.org/abs/1803.02837}{{\ttfamily arXiv:1803.02837 [hep-th]}}.

\bibitem{Ozsoy:2018flq}
O.~Ozsoy, S.~Parameswaran, G.~Tasinato, and I.~Zavala, ``{Mechanisms for
  Primordial Black Hole Production in String Theory},''
  \href{http://dx.doi.org/10.1088/1475-7516/2018/07/005}{{\em JCAP} {\bfseries
  1807} (2018) 005},
\href{http://arxiv.org/abs/1803.07626}{{\ttfamily arXiv:1803.07626 [hep-th]}}.

\bibitem{Mahbub:2019uhl}
R.~Mahbub, ``{Primordial black hole formation in inflationary
  $\alpha$-attractor models},''
  \href{http://dx.doi.org/10.1103/PhysRevD.101.023533}{{\em Phys. Rev. D}
  {\bfseries 101} no.~2, (2020) 023533},
  \href{http://arxiv.org/abs/1910.10602}{{\ttfamily arXiv:1910.10602
  [astro-ph.CO]}}.

\bibitem{Ballesteros:2020qam}
G.~Ballesteros, J.~Rey, M.~Taoso, and A.~Urbano, ``{Primordial black holes as
  dark matter and gravitational waves from single-field polynomial
  inflation},'' \href{http://dx.doi.org/10.1088/1475-7516/2020/07/025}{{\em
  JCAP} {\bfseries 07} (2020) 025},
  \href{http://arxiv.org/abs/2001.08220}{{\ttfamily arXiv:2001.08220
  [astro-ph.CO]}}.

\bibitem{Liu:2020oqe}
J.~Liu, Z.-K. Guo, and R.-G. Cai, ``{Analytical approximation of the scalar
  spectrum in the ultraslow-roll inflationary models},''
  \href{http://dx.doi.org/10.1103/PhysRevD.101.083535}{{\em Phys. Rev. D}
  {\bfseries 101} no.~8, (2020) 083535},
  \href{http://arxiv.org/abs/2003.02075}{{\ttfamily arXiv:2003.02075
  [astro-ph.CO]}}.

\bibitem{Kefala:2020xsx}
K.~Kefala, G.~P. Kodaxis, I.~D. Stamou, and N.~Tetradis, ``{Features of the
  inflaton potential and the power spectrum of cosmological perturbations},''
  \href{http://arxiv.org/abs/2010.12483}{{\ttfamily arXiv:2010.12483
  [astro-ph.CO]}}.

\bibitem{Carrilho:2019oqg}
P.~Carrilho, K.~A. Malik, and D.~J. Mulryne, ``{Dissecting the growth of the
  power spectrum for primordial black holes},''
  \href{http://dx.doi.org/10.1103/PhysRevD.100.103529}{{\em Phys. Rev. D}
  {\bfseries 100} no.~10, (2019) 103529},
  \href{http://arxiv.org/abs/1907.05237}{{\ttfamily arXiv:1907.05237
  [astro-ph.CO]}}.

\bibitem{Wands:1998yp}
D.~Wands, ``{Duality invariance of cosmological perturbation spectra},''
  \href{http://dx.doi.org/10.1103/PhysRevD.60.023507}{{\em Phys. Rev.}
  {\bfseries D60} (1999) 023507},
\href{http://arxiv.org/abs/gr-qc/9809062}{{\ttfamily arXiv:gr-qc/9809062
  [gr-qc]}}.

\bibitem{Kinney_2005}
W.~H. Kinney, ``Horizon crossing and inflation with large eta,''
  \href{http://dx.doi.org/10.1103/physrevd.72.023515}{{\em Physical Review D}
  {\bfseries 72} no.~2, (Jul, 2005) }.
  \url{http://dx.doi.org/10.1103/PhysRevD.72.023515}.

\bibitem{Tzirakis:2007bf}
K.~Tzirakis and W.~H. Kinney, ``{Inflation over the hill},''
  \href{http://dx.doi.org/10.1103/PhysRevD.75.123510}{{\em Phys. Rev.}
  {\bfseries D75} (2007) 123510},
\href{http://arxiv.org/abs/astro-ph/0701432}{{\ttfamily arXiv:astro-ph/0701432
  [astro-ph]}}.

\bibitem{Morse:2018kda}
M.~J.~P. Morse and W.~H. Kinney, ``{Large-$\eta$ constant-roll inflation is
  never an attractor},''
  \href{http://dx.doi.org/10.1103/PhysRevD.97.123519}{{\em Phys. Rev.}
  {\bfseries D97} no.~12, (2018) 123519},
\href{http://arxiv.org/abs/1804.01927}{{\ttfamily arXiv:1804.01927
  [astro-ph.CO]}}.

\bibitem{Atal:2018neu}
V.~Atal and C.~Germani, ``{The role of non-gaussianities in Primordial Black
  Hole formation},'' \href{http://dx.doi.org/10.1016/j.dark.2019.100275}{{\em
  Phys. Dark Univ.} {\bfseries 24} (2019) 100275},
\href{http://arxiv.org/abs/1811.07857}{{\ttfamily arXiv:1811.07857
  [astro-ph.CO]}}.

\bibitem{Chluba:2015bqa}
J.~Chluba, J.~Hamann, and S.~P. Patil, ``{Features and New Physical Scales in
  Primordial Observables: Theory and Observation},''
  \href{http://dx.doi.org/10.1142/S0218271815300232}{{\em Int. J. Mod. Phys.}
  {\bfseries D24} no.~10, (2015) 1530023},
\href{http://arxiv.org/abs/1505.01834}{{\ttfamily arXiv:1505.01834
  [astro-ph.CO]}}.

\bibitem{Chen:2005fe}
X.~Chen, ``{Running non-Gaussianities in DBI inflation},''
  \href{http://dx.doi.org/10.1103/PhysRevD.72.123518}{{\em Phys. Rev.}
  {\bfseries D72} (2005) 123518},
\href{http://arxiv.org/abs/astro-ph/0507053}{{\ttfamily arXiv:astro-ph/0507053
  [astro-ph]}}.

\bibitem{Byrnes:2009pe}
C.~T. Byrnes, S.~Nurmi, G.~Tasinato, and D.~Wands, ``{Scale dependence of local
  fNL},'' \href{http://dx.doi.org/10.1088/1475-7516/2010/02/034}{{\em JCAP}
  {\bfseries 1002} (2010) 034},
\href{http://arxiv.org/abs/0911.2780}{{\ttfamily arXiv:0911.2780
  [astro-ph.CO]}}.

\bibitem{Byrnes:2010ft}
C.~T. Byrnes, M.~Gerstenlauer, S.~Nurmi, G.~Tasinato, and D.~Wands,
  ``{Scale-dependent non-Gaussianity probes inflationary physics},''
  \href{http://dx.doi.org/10.1088/1475-7516/2010/10/004}{{\em JCAP} {\bfseries
  1010} (2010) 004},
\href{http://arxiv.org/abs/1007.4277}{{\ttfamily arXiv:1007.4277
  [astro-ph.CO]}}.

\bibitem{Passaglia:2018ixg}
S.~Passaglia, W.~Hu, and H.~Motohashi, ``{Primordial black holes and local
  non-Gaussianity in canonical inflation},''
  \href{http://dx.doi.org/10.1103/PhysRevD.99.043536}{{\em Phys. Rev. D}
  {\bfseries 99} no.~4, (2019) 043536},
  \href{http://arxiv.org/abs/1812.08243}{{\ttfamily arXiv:1812.08243
  [astro-ph.CO]}}.

\bibitem{Gao:2021vxb}
Q.~Gao, ``{Primordial black holes and secondary gravitational waves from
  chaotic inflation},'' \href{http://arxiv.org/abs/2102.07369}{{\ttfamily
  arXiv:2102.07369 [gr-qc]}}.

\bibitem{DeLuca:2021hcf}
V.~De~Luca, G.~Franciolini, and A.~Riotto, ``{Constraining the Initial
  Primordial Black Hole Clustering with CMB-distortion},''
\href{http://arxiv.org/abs/2103.16369}{{\ttfamily arXiv:2103.16369
  [astro-ph.CO]}}.

\bibitem{Unal:2020mts}
C.~Unal, E.~D. Kovetz, and S.~P. Patil, ``{Multi-messenger Probes of
  Inflationary Fluctuations and Primordial Black Holes},''
  \href{http://arxiv.org/abs/2008.11184}{{\ttfamily arXiv:2008.11184
  [astro-ph.CO]}}.

\bibitem{Garcia-Bellido:2017fdg}
J.~García-Bellido, ``{Massive Primordial Black Holes as Dark Matter and their
  detection with Gravitational Waves},''
  \href{http://dx.doi.org/10.1088/1742-6596/840/1/012032}{{\em J. Phys. Conf.
  Ser.} {\bfseries 840} no.~1, (2017) 012032},
\href{http://arxiv.org/abs/1702.08275}{{\ttfamily arXiv:1702.08275
  [astro-ph.CO]}}.

\bibitem{Garcia-Bellido:2017aan}
J.~Garcia-Bellido, M.~Peloso, and C.~Unal, ``{Gravitational Wave signatures of
  inflationary models from Primordial Black Hole Dark Matter},''
  \href{http://dx.doi.org/10.1088/1475-7516/2017/09/013}{{\em JCAP} {\bfseries
  09} (2017) 013}, \href{http://arxiv.org/abs/1707.02441}{{\ttfamily
  arXiv:1707.02441 [astro-ph.CO]}}.

\bibitem{Bravo:2020hde}
R.~Bravo and G.~A. Palma, ``{Unifying attractor and non-attractor models of
  inflation under a single soft theorem},''
\href{http://arxiv.org/abs/2009.03369}{{\ttfamily arXiv:2009.03369 [hep-th]}}.

\bibitem{Suyama:2021adn}
T.~Suyama, Y.~Tada, and M.~Yamaguchi, ``{Revisiting non-Gaussianity in
  non-attractor inflation models in the light of the cosmological soft
  theorem},''
\href{http://arxiv.org/abs/2101.10682}{{\ttfamily arXiv:2101.10682 [hep-th]}}.

\bibitem{Matarrese:2020why}
S.~Matarrese, L.~Pilo, and R.~Rollo, ``{Resilience of long modes in
  cosmological observables},''
  \href{http://dx.doi.org/10.1088/1475-7516/2021/01/062}{{\em JCAP} {\bfseries
  2101} (2021) 062},
\href{http://arxiv.org/abs/2007.08877}{{\ttfamily arXiv:2007.08877
  [astro-ph.CO]}}.

\bibitem{Taoso:2021uvl}
M.~Taoso and A.~Urbano, ``{Non-gaussianities for primordial black hole
  formation},''
\href{http://arxiv.org/abs/2102.03610}{{\ttfamily arXiv:2102.03610
  [astro-ph.CO]}}.

\bibitem{Cai:2017bxr}
Y.-F. Cai, X.~Chen, M.~H. Namjoo, M.~Sasaki, D.-G. Wang, and Z.~Wang,
  ``{Revisiting non-Gaussianity from non-attractor inflation models},''
\href{http://arxiv.org/abs/1712.09998}{{\ttfamily arXiv:1712.09998
  [astro-ph.CO]}}.

\bibitem{Namjoo:2012aa}
M.~H. Namjoo, H.~Firouzjahi, and M.~Sasaki, ``{Violation of non-Gaussianity
  consistency relation in a single field inflationary model},''
  \href{http://dx.doi.org/10.1209/0295-5075/101/39001}{{\em EPL} {\bfseries
  101} no.~3, (2013) 39001},
\href{http://arxiv.org/abs/1210.3692}{{\ttfamily arXiv:1210.3692
  [astro-ph.CO]}}.

\bibitem{Martin:2012pe}
J.~Martin, H.~Motohashi, and T.~Suyama, ``{Ultra Slow-Roll Inflation and the
  non-Gaussianity Consistency Relation},''
  \href{http://dx.doi.org/10.1103/PhysRevD.87.023514}{{\em Phys. Rev.}
  {\bfseries D87} no.~2, (2013) 023514},
\href{http://arxiv.org/abs/1211.0083}{{\ttfamily arXiv:1211.0083
  [astro-ph.CO]}}.

\bibitem{Chen:2013aj}
X.~Chen, H.~Firouzjahi, M.~H. Namjoo, and M.~Sasaki, ``{A Single Field
  Inflation Model with Large Local Non-Gaussianity},''
  \href{http://dx.doi.org/10.1209/0295-5075/102/59001}{{\em EPL} {\bfseries
  102} no.~5, (2013) 59001},
\href{http://arxiv.org/abs/1301.5699}{{\ttfamily arXiv:1301.5699 [hep-th]}}.

\bibitem{Chen:2013eea}
X.~Chen, H.~Firouzjahi, E.~Komatsu, M.~H. Namjoo, and M.~Sasaki, ``{In-in and
  $\delta N$ calculations of the bispectrum from non-attractor single-field
  inflation},'' \href{http://dx.doi.org/10.1088/1475-7516/2013/12/039}{{\em
  JCAP} {\bfseries 1312} (2013) 039},
\href{http://arxiv.org/abs/1308.5341}{{\ttfamily arXiv:1308.5341
  [astro-ph.CO]}}.

\bibitem{Finelli:2017fml}
B.~Finelli, G.~Goon, E.~Pajer, and L.~Santoni, ``{Soft Theorems For
  Shift-Symmetric Cosmologies},''
  \href{http://dx.doi.org/10.1103/PhysRevD.97.063531}{{\em Phys. Rev. D}
  {\bfseries 97} no.~6, (2018) 063531},
  \href{http://arxiv.org/abs/1711.03737}{{\ttfamily arXiv:1711.03737
  [hep-th]}}.

\bibitem{Sunyaev:1970er}
R.~A. Sunyaev and Y.~B. Zeldovich, ``{The Interaction of matter and radiation
  in the hot model of the universe},'' {\em Astrophys. Space Sci.} {\bfseries
  7} (1970) 20--30.

\bibitem{Sunyaev:1970er2}
R.~A. Sunyaev and Y.~B. Zeldovich, ``{Small scale entropy and adiabatic density
  perturbations: Antimatter in the Universe},'' {\em Astrophys. Space Sci.}
  {\bfseries 9} 368--382.

\bibitem{Hu:1994bz}
W.~Hu, D.~Scott, and J.~Silk, ``{Power spectrum constraints from spectral
  distortions in the cosmic microwave background},''
  \href{http://dx.doi.org/10.1086/187424}{{\em Astrophys. J. Lett.} {\bfseries
  430} (1994) L5--L8}, \href{http://arxiv.org/abs/astro-ph/9402045}{{\ttfamily
  arXiv:astro-ph/9402045}}.

\bibitem{Khatri:2011aj}
R.~Khatri, R.~A. Sunyaev, and J.~Chluba, ``{Does Bose-Einstein condensation of
  CMB photons cancel \textbackslash{}mu distortions created by dissipation of
  sound waves in the early Universe?},''
  \href{http://dx.doi.org/10.1051/0004-6361/201118194}{{\em Astron. Astrophys.}
  {\bfseries 540} (2012) A124},
  \href{http://arxiv.org/abs/1110.0475}{{\ttfamily arXiv:1110.0475
  [astro-ph.CO]}}.

\bibitem{Chluba:2011hw}
J.~Chluba and R.~A. Sunyaev, ``{The evolution of CMB spectral distortions in
  the early Universe},''
  \href{http://dx.doi.org/10.1111/j.1365-2966.2011.19786.x}{{\em Mon. Not. Roy.
  Astron. Soc.} {\bfseries 419} (2012) 1294--1314},
  \href{http://arxiv.org/abs/1109.6552}{{\ttfamily arXiv:1109.6552
  [astro-ph.CO]}}.

\bibitem{Chluba:2012gq}
J.~Chluba, R.~Khatri, and R.~A. Sunyaev, ``{CMB at 2x2 order: The dissipation
  of primordial acoustic waves and the observable part of the associated energy
  release},'' \href{http://dx.doi.org/10.1111/j.1365-2966.2012.21474.x}{{\em
  Mon. Not. Roy. Astron. Soc.} {\bfseries 425} (2012) 1129--1169},
  \href{http://arxiv.org/abs/1202.0057}{{\ttfamily arXiv:1202.0057
  [astro-ph.CO]}}.

\bibitem{Nakama:2017ohe}
T.~Nakama, J.~Chluba, and M.~Kamionkowski, ``{Shedding light on the small-scale
  crisis with CMB spectral distortions},''
  \href{http://dx.doi.org/10.1103/PhysRevD.95.121302}{{\em Phys. Rev. D}
  {\bfseries 95} no.~12, (2017) 121302},
  \href{http://arxiv.org/abs/1703.10559}{{\ttfamily arXiv:1703.10559
  [astro-ph.CO]}}.

\bibitem{Ganc:2012ae}
J.~Ganc and E.~Komatsu, ``{Scale-dependent bias of galaxies and mu-type
  distortion of the cosmic microwave background spectrum from single-field
  inflation with a modified initial state},''
  \href{http://dx.doi.org/10.1103/PhysRevD.86.023518}{{\em Phys. Rev. D}
  {\bfseries 86} (2012) 023518},
  \href{http://arxiv.org/abs/1204.4241}{{\ttfamily arXiv:1204.4241
  [astro-ph.CO]}}.

\bibitem{Komatsu:2001rj}
E.~Komatsu and D.~N. Spergel, ``{Acoustic signatures in the primary microwave
  background bispectrum},''
  \href{http://dx.doi.org/10.1103/PhysRevD.63.063002}{{\em Phys. Rev. D}
  {\bfseries 63} (2001) 063002},
  \href{http://arxiv.org/abs/astro-ph/0005036}{{\ttfamily
  arXiv:astro-ph/0005036}}.

\bibitem{Biagetti:2013sr}
M.~Biagetti, H.~Perrier, A.~Riotto, and V.~Desjacques, ``{Testing the running
  of non-Gaussianity through the CMB $\mu$-distortion and the halo bias},''
  \href{http://dx.doi.org/10.1103/PhysRevD.87.063521}{{\em Phys. Rev. D}
  {\bfseries 87} (2013) 063521},
  \href{http://arxiv.org/abs/1301.2771}{{\ttfamily arXiv:1301.2771
  [astro-ph.CO]}}.

\bibitem{Chluba:2016aln}
J.~Chluba, E.~Dimastrogiovanni, M.~A. Amin, and M.~Kamionkowski, ``{Evolution
  of CMB spectral distortion anisotropies and tests of primordial
  non-Gaussianity},'' \href{http://dx.doi.org/10.1093/mnras/stw3230}{{\em Mon.
  Not. Roy. Astron. Soc.} {\bfseries 466} no.~2, (2017) 2390--2401},
  \href{http://arxiv.org/abs/1610.08711}{{\ttfamily arXiv:1610.08711
  [astro-ph.CO]}}.

\bibitem{Cabass:2018jgj}
G.~Cabass, E.~Pajer, and D.~van~der Woude, ``{Spectral distortion anisotropies
  from single-field inflation},''
  \href{http://dx.doi.org/10.1088/1475-7516/2018/08/050}{{\em JCAP} {\bfseries
  08} (2018) 050}, \href{http://arxiv.org/abs/1805.08775}{{\ttfamily
  arXiv:1805.08775 [astro-ph.CO]}}.

\bibitem{Kogut:2011xw}
A.~Kogut {\em et~al.}, ``{The Primordial Inflation Explorer (PIXIE): A Nulling
  Polarimeter for Cosmic Microwave Background Observations},''
  \href{http://dx.doi.org/10.1088/1475-7516/2011/07/025}{{\em JCAP} {\bfseries
  07} (2011) 025}, \href{http://arxiv.org/abs/1105.2044}{{\ttfamily
  arXiv:1105.2044 [astro-ph.CO]}}.

\bibitem{Andre:2013nfa}
{\bfseries PRISM} Collaboration, P.~Andr\'e {\em et~al.}, ``{PRISM (Polarized
  Radiation Imaging and Spectroscopy Mission): An Extended White Paper},''
  \href{http://dx.doi.org/10.1088/1475-7516/2014/02/006}{{\em JCAP} {\bfseries
  02} (2014) 006}, \href{http://arxiv.org/abs/1310.1554}{{\ttfamily
  arXiv:1310.1554 [astro-ph.CO]}}.

\bibitem{Akrami:2019izv}
{\bfseries Planck} Collaboration, Y.~Akrami {\em et~al.}, ``{Planck 2018
  results. IX. Constraints on primordial non-Gaussianity},''
  \href{http://dx.doi.org/10.1051/0004-6361/201935891}{{\em Astron. Astrophys.}
  {\bfseries 641} (2020) A9}, \href{http://arxiv.org/abs/1905.05697}{{\ttfamily
  arXiv:1905.05697 [astro-ph.CO]}}.

\bibitem{Garcia:2020mwi}
M.~A.~G. Garcia, M.~A. Amin, and D.~Green, ``{Curvature Perturbations From
  Stochastic Particle Production During Inflation},''
  \href{http://dx.doi.org/10.1088/1475-7516/2020/06/039}{{\em JCAP} {\bfseries
  06} (2020) 039}, \href{http://arxiv.org/abs/2001.09158}{{\ttfamily
  arXiv:2001.09158 [astro-ph.CO]}}.

\bibitem{Palma:2020ejf}
G.~A. Palma, S.~Sypsas, and C.~Zenteno, ``{Seeding primordial black holes in
  multifield inflation},''
  \href{http://dx.doi.org/10.1103/PhysRevLett.125.121301}{{\em Phys. Rev.
  Lett.} {\bfseries 125} no.~12, (2020) 121301},
  \href{http://arxiv.org/abs/2004.06106}{{\ttfamily arXiv:2004.06106
  [astro-ph.CO]}}.

\bibitem{Fumagalli:2020adf}
J.~Fumagalli, S.~Renaux-Petel, J.~W. Ronayne, and L.~T. Witkowski, ``{Turning
  in the landscape: a new mechanism for generating Primordial Black Holes},''
  \href{http://arxiv.org/abs/2004.08369}{{\ttfamily arXiv:2004.08369
  [hep-th]}}.

\bibitem{Fumagalli:2020nvq}
J.~Fumagalli, S.~Renaux-Petel, and L.~T. Witkowski, ``{Oscillations in the
  stochastic gravitational wave background from sharp features and particle
  production during inflation},''
  \href{http://arxiv.org/abs/2012.02761}{{\ttfamily arXiv:2012.02761
  [astro-ph.CO]}}.

\bibitem{Braglia:2020taf}
M.~Braglia, X.~Chen, and D.~Kumar~Hazra, ``{Probing Primordial Features with
  the Stochastic Gravitational Wave Background},''
  \href{http://dx.doi.org/10.1088/1475-7516/2021/03/005}{{\em JCAP} {\bfseries
  03} (2021) 005}, \href{http://arxiv.org/abs/2012.05821}{{\ttfamily
  arXiv:2012.05821 [astro-ph.CO]}}.

\bibitem{Garcia-Bellido:2016dkw}
J.~Garcia-Bellido, M.~Peloso, and C.~Unal, ``{Gravitational waves at
  interferometer scales and primordial black holes in axion inflation},''
  \href{http://dx.doi.org/10.1088/1475-7516/2016/12/031}{{\em JCAP} {\bfseries
  1612} no.~12, (2016) 031},
\href{http://arxiv.org/abs/1610.03763}{{\ttfamily arXiv:1610.03763
  [astro-ph.CO]}}.

\bibitem{Ozsoy:2020ccy}
O.~\"Ozsoy, ``{Synthetic Gravitational Waves from a Rolling Axion Monodromy},''
  \href{http://dx.doi.org/10.1088/1475-7516/2021/04/040}{{\em JCAP} {\bfseries
  04} (2021) 040}, \href{http://arxiv.org/abs/2005.10280}{{\ttfamily
  arXiv:2005.10280 [astro-ph.CO]}}.

\bibitem{Ozsoy:2020kat}
O.~\"Ozsoy and Z.~Lalak, ``{Primordial black holes as dark matter and
  gravitational waves from bumpy axion inflation},''
  \href{http://dx.doi.org/10.1088/1475-7516/2021/01/040}{{\em JCAP} {\bfseries
  01} (2021) 040}, \href{http://arxiv.org/abs/2008.07549}{{\ttfamily
  arXiv:2008.07549 [astro-ph.CO]}}.

\end{thebibliography}\endgroup

\end{document}